\documentclass[twocolumn]{aastex631}

\usepackage{amsmath}
\usepackage{enumitem, amssymb}
\usepackage[thinc]{esdiff} 
\usepackage{pifont}
\usepackage{xcolor}

\newlist{todolist}{itemize}{2}
\setlist[todolist]{label=$\square$}
\newcommand{\tr}[1]{\textrm{#1}}

\newcommand{\response}[1]{#1}

\newcommand{\snr}{\tr{S}/\tr{N}}
\newcommand{\sigdat}{\snr_\rho}
\newcommand{\sigsig}{\snr_\tr{s}}
\newcommand{\lnlike}{\snr_\Lambda}
\newcommand{\rchisq}{\chi_\nu^2}

\newcommand{\bfthresh}{\overline{\bfact}}

\newcommand{\normal}[2]{\mathcal{N}(#1, #2)}
\newcommand{\uniform}[2]{\mathcal{U}(#1, #2)}

\newcommand{\mmbamp}{\mu}
\newcommand{\mmbplaw}{\alpha_\mu}
\newcommand{\mmbscatter}{\epsilon_\mu}

\newcommand{\thard}{\tau_f}
\newcommand{\hardrchar}{a_c}
\newcommand{\hardainit}{a_\tr{init}}

\newcommand{\hardnuinner}{\nu_\tr{inner}}
\newcommand{\hardnuouter}{\nu_\tr{outer}}

\newcommand{\holodeck}{\texttt{holodeck}}
\newcommand{\classicuniform}{\texttt{Classic\_Uniform\_Phenom}}
\newcommand{\asa}{\texttt{Astro\_Strong\_All}}
\newcommand{\qcw}{\texttt{QuickCW}}

\newcommand{\mbh}{M_\tr{BH}}
\newcommand{\mbulge}{M_\tr{bulge}}
\newcommand{\mmbulge}{{\mbh\tr{--}\mbulge}}

\newcommand{\gsmffunc}{\Psi}
\newcommand{\gsmfmasszero}{m_{\psi,z_0}}

\newcommand{\tgal}{T_\tr{gal-gal}}
\newcommand{\gmrnormlog}{R_{0}}
\newcommand{\gmrnormz}{\eta_{r}}
\newcommand{\gmrmalphao}{\alpha_{r,0}}
\newcommand{\gmrmalphaz}{\alpha_{r,z}}
\newcommand{\gmrmdeltao}{\delta_{r,0}}
\newcommand{\gmrmdeltaz}{\delta_{r,z}}
\newcommand{\gmrqgammao}{\beta_{r,0}}
\newcommand{\gmrqgammaz}{\beta_{r,z}}
\newcommand{\gmrqgammam}{\gamma_{r}}

\newcommand{\bffbulge}{f_b}
\newcommand{\bfmchar}{m_{b,c}}
\newcommand{\bfflo}{f_{b,l}}
\newcommand{\bffhi}{f_{b,h}}
\newcommand{\bfwidth}{{k_b}}

\newcommand{\msol}{\tr{M}_{\odot}}

\newcommand{\yr}{\tr{yr}}        

\newcommand{\ndens}{\eta}
\newcommand{\ndensgalgal}{\eta_\tr{gal-gal}}

\newcommand{\mstartot}{M_{\star}}
\newcommand{\mstar}{m_{\star1}}

\newcommand{\qstar}{q_\star}

\newcommand{\mchirp}{\mathcal{M}_\tr{c}}     
\newcommand{\dl}{d_l}   

\newcommand{\hc}{h_\tr{c}}
\newcommand{\hs}{h_{0}}

\newcommand{\fo}[1][]{
    \ifthenelse{\equal{#1}{}}{
        f_\tr{o}
    }{
        {f_{\tr{o}, #1}}
    }
}

\newcommand{\absdeltafo}{|\delta (\fo)|}
\newcommand{\logfo}[1][]{
    \ifthenelse{\equal{#1}{}}{
        \log_{10}f_\tr{o}
    }{
        \log_{10}{f_{\tr{o}, #1}}
    }
}
\newcommand{\loghs}{\log_{10}\hs}
\newcommand{\logmc}{\log_{10}\mc}
\newcommand{\mc}{\mathcal{M}_\tr{c}}
\newcommand{\gwbamp}{A_\tr{GWB}}
\newcommand{\gwbgamma}{\gamma_\tr{GWB}}

\newcommand{\lr}[2][]{
    \ifthenelse{\equal{#1}{}}{
        {\left(#2\right)}
    }{
        {\left(#2\right)}^{#1}
    }
}

\newcommand{\lrs}[2][]{
    \ifthenelse{\equal{#1}{}}{
        {\left[#2\right]}
    }{
        {\left[#2\right]}^{#1}
    }
}

\newcommand{\scale}[3][]{
    \ifthenelse{\equal{#1}{}}{
        \lr{ \frac{#2}{#3} }
    }{
        {\lr[#1]{ \frac{#2}{#3} }}
    }
}

\newcommand{\innerprod}[2]{( #1 | #2 )}
\newcommand{\likelihood}[2]{\mathcal{L}(#1 | #2 )}
\newcommand{\signal}{\boldsymbol{s}}
\newcommand{\data}{\boldsymbol{d}}
\newcommand{\commonpars}{\vec{\boldsymbol{\lambda}}}
\newcommand{\model}{\mathcal{H}}
\newcommand{\bfact}{\mathcal{B}}
\newcommand{\ciwidth}[1]{\Delta_{68}(#1)}

\newcommand{\figref}[1]{Fig.~\ref{#1}}
\newcommand{\secref}[1]{\textsection\ref{#1}}

\newcommand{\tabref}[1]{{Table~\ref{#1}}}

\usepackage{amsmath}
\usepackage[thinc]{esdiff} 
\usepackage{xifthen}
\usepackage{txfonts}
\usepackage{xfrac}
\usepackage{enumitem}
\usepackage{hyperref}
\usepackage{makecell}

\begin{document}

\title{Characterizing Continuous Gravitational Waves from Supermassive Black Hole Binaries in Realistic Pulsar Timing Array Data}

\author[0000-0002-8857-613X]{Emiko C. Gardiner}
\affiliation{Department of Astronomy, University of California, Berkeley 501 Campbell Hall \#3411 Berkeley, CA 94720, USA}

\author[0000-0003-0909-5563]{Bence B\'ecsy}
\affiliation{Department of Physics, Oregon State University, Corvallis, OR 97331, USA}
\affiliation{Institute for Gravitational Wave Astronomy and School of Physics and Astronomy, University of Birmingham, Edgbaston, Birmingham B15 2TT, UK}

\author[0000-0002-6625-6450]{Luke Zoltan Kelley}
\affiliation{Department of Astronomy, University of California, Berkeley 501 Campbell Hall \#3411 Berkeley, CA 94720, USA}

\author[0000-0002-7435-0869]{Neil J. Cornish}
\affiliation{Department of Physics, Montana State University, Bozeman, MT 59717, USA}


\begin{abstract}
Pulsar timing arrays recently found evidence for a gravitational wave background (GWB), likely the stochastic overlap of GWs from many supermassive black hole binaries. 
Anticipating a continuous gravitational wave (CW) detection from a single binary soon to follow, we examine how well current Bayesian methods can detect CWs and characterize their binary properties by
modeling the response of the NANOGrav 15-year pulsar timing array to simulated binary populations. 
We run Markov Chain Monte Carlo searches for CWs in these datasets and compare them to quicker detection statistics including the optimal signal-to-noise ratio (S/N), matched filter detection statistic, and reduced log-likelihood ratio between the signal and noise models calculated at the injected parameters.
The latter is the best proxy for Bayesian detection fractions, corresponding to a 50\% detection fraction (by Bayes factors $>10$ favoring a CW detection over noise-only model) at an $\snr = 4.6$. 
Source confusion between the GWB and a CW, or between multiple CWs, can cause false detections and unexpected dismissals. 
53\% of realistic binary populations consistent with the recently observed GWB have successful CW detections.
82\% of these CWs are in the 4th or 5th frequency bin of the 16.03 yr dataset (6.9 and 10.8~nHz), with 95th percentile regions spanning 
$4$--$12~\tr{nHz}$ frequencies,
\response{0.}$7$--$20~\times~10^9~\msol$ chirp masses, 
$60~\tr{Mpc}$--$8~\tr{Gpc}$ luminosity distances, and $18~\deg^2$--$13,000~\deg^2$ 68\% confidence localization areas. 
These successful detections often poorly recover the chirp mass, with only 29\% identifying the chirp mass accurately to within 1 dex with a 68\% posterior width also narrower than 1 dex.
\end{abstract}


\section{Introduction} \label{sec:intro}
Black hole binaries in stable orbits emit near-monochromatic gravitational waves (GWs), known as continuous waves (CWs). 
According to hierarchical structure formation, supermassive black hole binaries (SMBHBs) result from galaxy mergers and emit these CWs at nanohertz frequencies when they reach sufficiently small ($\sim\!\!\tr{mpc}$) separations \citep{volonteri+2003, sesana+2009}. 
Pulsar timing arrays (PTAs) can detect these nanohertz gravitational waves by monitoring the times of arrival (TOAs) of radio pulses from millisecond pulsars across our Galaxy for years to decades \response{
\citep{sazhin_1978, detweiler_1979, foster+backer1990, burkespolaor+2019, taylor2021}}.  
While an individual continuous wave has not yet been detected \response{\citep{ppta+2010, ng+2014, ppta+2014, epta+2016, ng+2019, epta+23_individuals}}, multiple PTAs have found evidence of a gravitational wave background \response{\citep[GWB;][]{ng+23_gwb, epta+23_gwb, ppta+23_gwb, cpta+23, mpta+2025}} consistent with the stochastic red noise of many superimposed continuous waves \citep{ng+23_astro}. 
In the likely scenario that SMBHBs produce the GWB, a CW detection is predicted to follow \citep{rosado+2015}, potentially within a few years \citep{kelley+2018}. 
Various cosmological processes could also contribute to the GWB \citep{ng+23_newphysics, epta+23_physics}. 
A CW detection would definitively confirm the SMBHB model, while a continued lack of CW detection could support alternative models.

In preparation for near-future CW detections, we analyze the efficacy of current detection methods applied to GW signals from realistic populations of simulated SMBHBs, in the form of realistic TOA datasets. 
Specifically, we tailor this analysis to the pulsar observing times, TOA errors, and sky locations in the North American Nanohertz Observatory for Gravitational Waves (NANOGrav) 15-year dataset \citep[henceforth NG15]{ng+23_timing} and the Markov Chain Monte Carlo (MCMC) pipeline \citep{becsy+2022_quickcw} used in the corresponding CW search \citep{ng+23_individuals}.
This pipeline treats the background as a red noise process without including spatial correlations. Fast analyses capable of including correlations are still being developed \citep[e.g.,][]{gunderson+cornish2025}, for which this study should be repeated to understand the impact of correlations in the background. 
Trans-dimensionality allowing for multiple CWs will be another key improvement \citep[e.g.,][]{becys+cornish2020, taylor+2020} to incorporate with faster methods in the future.

We quantify how well the pipeline detects CWs and measures their source properties across a wide variety of binary populations. 
Previous studies have taken a Fisher matrix approach to examine the measurement precision of these binary properties \citep{sesana+vecchio2010, Liu+2023}.
However, this method is less reliable in the low-to-mid signal-to-noise ratio ($\snr$) regime, where the parameter posterior distributions are non-Gaussian. 
Generating full Bayesian posteriors with the MCMC pipeline allows us to better characterize the measurement precision without assuming the shape of the posteriors. 

To characterize the distribution of detectable CW properties, we narrow this analysis to a subset of samples consistent with current electromagnetic (EM) and GWB observations. 
This builds on the predictions for binary properties and CW detectability in \citet{gardiner+2024} and \citet{cella+2025} by incorporating astrophysical and GWB constraints into fully realistic TOA datasets. 
Similar realistic simulations based on Illustris populations were explored in \citet{becsy+2022_realistic} and \citet{becsy+2023_howtodetect}, but these populations tend to produce GWBs at lower amplitudes than recent observations. We follow up on these works using semi-analytic binary population models to produce populations more consistent with observations and focus on CWs with a full Bayesian pipeline instead of simplified detection proxies.

CW detections and upper limits also encode information about SMBHB populations, their merger histories,  galaxy interactions, and binary evolution \citep{sesana+2009, kelley+2018, gardiner+2024}. 
Many model components are degenerate in how they impact the GWB, with variations that increase the characteristic number of binaries and peak mass of binaries similarly increasing the GWB amplitude. 
Meanwhile, CW detections are favored by models that increase the presence of high-mass outliers as opposed to the total number of sources \citep{satopolito+2025_distribution}. 
For example, a larger binary number density, and fast binary evolution, boost the GWB amplitude more than CW detectability, while greater scatter in $M_\tr{BH}$--$M_\tr{bulge}$ or $M_\tr{BH}$--$\sigma$ preferentially raise the probability of CW detection \citep{gardiner+2024}.

Interpreting simulated populations of SMBHBs is a crucial step in extracting this information. 
To understand exactly how each population would be observed, one can analyze the corresponding TOAs the same way as real data: defining a likelihood and constructing Bayesian posteriors by MCMC sampling. 
This method is robust, but computationally expensive, especially for CWs which have orders of magnitude more parameters than simpler GWB-only models. 
Applying such a process to hundreds of thousands of realizations, as is needed to place constraints on a large model parameter space, is unrealistic.

Instead, CW detectability in simulated populations has historically been assessed by comparing CW strain amplitudes to sensitivity curves \citep{larson+2000, ppta+2010, hazboun+2019} or calculating a CW detection statistic \citep{anholm+2008}.
Typical forms of detection statistics compare the likelihood ratio of a signal and non-signal model to construct an $\snr$.
For example, the $\mathcal{F}$-statistic is the reduced log-likelihood function for the maximum-likelihood estimators of each model parameter, originally developed by \citet{jaranowski+2009} to describe the detectability of continuous waves from neutron stars by the Laser Interferometer Gravitational-wave Observatory. \citet{cutler+schutz2005} generalized the $\mathcal{F}$-statistic to multidetector networks, \citet{cornish+porter2007} applied it to the Laser Interferometer Space Antenna, and \citet{sesana+vecchio2010} and \citet{babak+sesana2012} applied it to PTAs. 
\citet{ellis+2012} demonstrated that for a large number of pulsars, the pulsar terms cancel out and can be dropped in the log-likelihood ratio calculation, resulting in the Earth-term-only $\mathcal{F}_e$-statistic. 
\citet{rosado+2015} used this $\mathcal{F}_e$-statistic to compare CW and GWB detection probabilities over time for ideal uniform-duration simulated PTAs and \citet{taylor+2016} generalized the $\mathcal{F}_e$-statistic to eccentric systems.
\citet{taylor+2014} introduced variations of the $\mathcal{F}$-statistic that keep the pulsar terms, since they are important for measuring mass, distance, and sky location, but marginalize over the pulsar-phase variables.

We test several of the $\snr$s used throughout these detection statistic formalisms including the optimal matched filter $\snr$ ($\sigsig$) which only accounts for the strength of the injected signal, the maximum-likelihood matched filter detection statistic ($\sigdat$) which also incorporates the noise component of the data, and the reduced log-likelihood ratio corresponding to the $\mathcal{F}$-statistic ($\lnlike$) calculated at the injected parameters.
These $\snr$s can all be calculated directly from the TOAs of a signal and the signal model.
However, we find that none map directly to the output of full Bayesian detection analyses. 
We use our ensemble of SMBHB populations to establish the best mapping between TOA-based $\snr$s 
and MCMC-established detection fractions. 
This relation makes it possible to rapidly explore realistic CW detectability across broad parameter spaces, which we will present in a follow-up paper.

The remainder of the paper is organized as follows. 
In \secref{sec:methods_pops}--\secref{sec:methods_snrs} we describe how we generate the SMBHB populations, create their corresponding TOAs, and calculated TOA-based SNRs.
\secref{sec:methods_quickcw} covers the realistic MCMC detection pipeline process and interpretation and \secref{sec:methods_map} the relationship between SNRs and Bayes Factor-based detection fractions. 
Then, we present the best threshold Bayes factor and corresponding detection rates in \secref{sec:results_quickcw}, 
the $\snr$ to detection fraction mapping in \secref{sec:results_map}, 
the binary property recovery in \secref{sec:results_properties}, 
and a realistic GWB-conditioned population analysis in \secref{sec:results_realpop}. 
Finally, we discuss ways to improve CW detection and the implications and future applications of this work in \secref{sec:discussion} and summarize the key conclusions in \secref{sec:conclusion}.


\section{Methods} \label{sec:methods}


\subsection{Simulating SMBHB Populations} \label{sec:methods_pops}

We aim to characterize how well current search methods can detect and recover the properties of CWs, under a wide variety of astrophysical scenarios.
To do so, we begin by generating simulated populations of SMBHBs with \texttt{holodeck} \citep{holodeck}, taking the semi-analytic model (SAM) approach as detailed in \citet{ng+23_astro}.
We use a variety of SAMs, all of which begin by calculating a distribution of galaxies over mass and initial redshift  $\Psi(m_{*1},z)$ following a single- or double-Schechter galaxy stellar mass function (GSMF).
We translate the initial distribution of galaxies into post-merger galaxies using power-law functions of galaxy stellar mass $\mstar$, stellar mass ratio $\qstar$, and initial redshift $z'$ for the galaxy pair fraction $P$ and galaxy merger time $T_\tr{gal-gal}$, 
\begin{equation}
\label{eq:ndens_galgal}
        \diffp{\ndensgalgal}{{\mstar}{\qstar}{z'}} = \frac{\gsmffunc(\mstar,z')}{\mstar \ln\! \lr{10}} \, \frac{P(\mstar,\qstar,z')}{\tgal(\mstar,\qstar,z')} \diffp{t}{{z'}}.
\end{equation}
or using a multiple-power-law galaxy merger rate $R$,
\begin{equation}
\label{eq:ndens_galgal_gmr}
        \diffp{\ndensgalgal}{{\mstar}{\qstar}{z'}} = \frac{\gsmffunc(\mstar,z')}{\mstar \ln\! \lr{10}} \, R(\mstartot,\qstar,z') \diffp{t}{{z'}}.
\end{equation}
where $\mstar$ represents the more massive galaxy, $\mstartot$ is the total mass $\mstar + m_{\star2}$, and the mass ratio is defined such that $\qstar= \mstar/m_{\star2} \leq 1$.
Then, we identify the SMBHB pairs associated with these post-merger host galaxies by a relation between black hole mass ($M=m_1+m_2$, $q=m_2/m_1 \leq 1$) and host galaxy mass,
\begin{equation}
        \diffp{\ndens}{{M}{q}{z'}} = \diffp{\ndensgalgal}{{\mstar}{\qstar}{z'}} \: \diffp{\mstar}{M} \: \diffp{\qstar}{q}.
\end{equation}

We evolve the binaries in time and binary separation using a phenomenological double power-law function encapsulating both large and small-scale environmental effects, in addition to GW hardening, to find the number density of binaries emitting GWs in the PTA frequency band as a function of total black hole mass $M$, mass ratio $q$, and final redshift $z$.
This leaves the comoving volumetric number density of SMBHBs $\frac{\partial^3 \eta}{\partial M \partial q \partial z}$, which is converted to an expectation value for the number of binaries at each $M$, $q$, $z$, and rest frame orbital frequent $f_p$ as
    \begin{equation}
        \label{eq:number_density_to_number_frequency}
        \diffp{N}{{M} {q} {z} {\ln f_p}} = \diffp{\ndens}{{M} {q} {z}} \diffp{t}{{\ln f_p}} \diffp{z}{{t}} \diffp{V_c}{{z}}. 
    \end{equation}
Finally, to generate discrete binary populations representing individual universe realizations, we randomly sample from a Poisson distribution $\mathcal{P}$ around the expectation value integrated over each parameter bin, to find
\begin{eqnarray}
    \label{eq:number_sampling}
    N(M,\!q,\!z,\!f) = \quad \quad \quad \quad \quad & \nonumber \\
     \mathcal{P}\,\Big( \diffp{N}{{M'} {q'} {z'}{\!\ln f_p'}} &
    \Delta M' \! \Delta q'\!\Delta z' \! \Delta \!\ln f'\Big)\Bigg|_{M, q, z, f_p}.
\end{eqnarray}

\subsection{Population Sample Selection} \label{sec:methods_samples}

\begin{deluxetable}{cccc}
\label{tab:subsets}
\tablecaption{Binary Population Sample Subsets}
\tablehead{
\colhead{Subset} & \colhead{Parameter Space} & \thead{Varied \\ Parameters} & \colhead{Samples} 
} 
\startdata
Low $\snr$ & $\classicuniform$ & $\thard$ & 400\\
High $\snr$ & $\asa$ & $\gsmfmasszero, \mmbscatter$ & 400 \\
Realistic & $\asa$ & all & 1200\\ 
\enddata
\end{deluxetable}

The power-law prescriptions for galaxy merger rate, BH--galaxy relation, and binary evolution introduce a number of free parameters, many of which are partially degenerate. 
We vary the parameters, or subsets of the parameters, in these SAMs to generate binary populations from many astrophysical scenarios in three subsets, outlined in \tabref{tab:subsets}. 
The columns of \tabref{tab:subsets} list, from left to right, the purpose of each subset, the $\holodeck$ SAM parameter space used, the parameters within that library that are varied, and the number of samples generated.

For the first subset, we draw from the broad, uniform parameter space used in \citet{ng+23_astro} and \citet{gardiner+2024}, titled $\classicuniform$ in \holodeck. 
By using broad, uninformative priors, this parameter space allowed all constraints in \citet{ng+23_astro} to be based on the gravitational wave measurement. 
We use the fiducial values for each parameter listed in \citet{ng+23_astro} Table B1, or the midpoint of their uniform prior range when no fiducial value is given, for all parameters except the binary lifetime. 
In the interest of creating a wide variety of populations and CW detectability, we vary the lifetime, $\thard$, the parameter that impacts the GWB and single-source outliers most distinctly \citep{gardiner+2024}. 
Most of the resulting $\classicuniform$ realizations still tend to have GWB amplitudes below the NG15 measurement and low CW $\snr$s.

To create a mapping between $\snr$s and Bayesian detectability, numerous high $\snr$ samples are also necessary. 
Thus, we design our second subset to generate higher $\snr$ samples. 
We adopt the more astrophysically motivated $\holodeck$ parameter space titled \texttt{Astro\_Strong\_All}, which is detailed in Appendix \ref{sec:appendix_pspace}.
For simplicity, all parameters are fixed to their fiducial values in \tabref{tab:asa_pspace} except for the scatter in the $M_\tr{BH}$--$M_\tr{bulge}$ relation $\mmbscatter$ and the characteristic mass of the GSMF $M_{\psi,z_0}$, which are varied between their fiducial values and maximum values of $\mmbscatter=0.8$ and $\gsmfmasszero=12$ to produce high $\snr$s. 

Finally, we generate the third subset to be as realistic as possible. 
Like the second subset, we use the astrophysically motivated $\asa$ parameter space, but now we marginalize over all of the parameters, using Latin-hypercube sampling from the parameter distributions in \tabref{tab:asa_pspace} to create 1000 distinct models. 
The binary evolution parameters $\tau_f$ and $\nu_\tr{inner}$ are drawn from uniform distributions since they lack well-informed EM priors, and the rest are drawn from normal distributions based on literature measurements and uncertainties (see Appendix \ref{sec:appendix_pspace}). 
For each model, we Poisson-sample 500 realizations, producing $1000 \times 500$ astrophysically realistic universes. 

Then, we incorporate GWB information by comparing each realization's characteristic strain spectrum to the \citet{ng+23_gwb} measurement, which found a Hellings and Downs (HD) correlated power-law spectrum with an amplitude $A_\tr{HD}=6.4^{+4.2}_{-0.6} \times 10^{-15}$ at reference frequency $1/(10~\yr)$ and spectral index $\gamma_\tr{HD}=3.2^{+0.6}_{-0.6}$.
We fit a power-law background to the first five frequency bins of each realization to get amplitude $A_{10~\yr}$ and spectral index $\gamma$ at the same reference frequency of $1/(10~\yr)$. Then, each sample is assigned a weight $w$ based on an uncorrelated normal distribution around the \citet{ng+23_gwb} $\log_{10}(A_\tr{HD})$ and $\gamma_\tr{HD}$, 
\begin{multline}\label{eq:gwb_weight}
    w(\log_{10}(A_{10~\yr}), \gamma) \\
    = \mathcal{N}(\log_{10}(A_{10~\yr}); -14.1, 0.1) \times \mathcal{N}(\gamma; 3.2, 0.3)
\end{multline} 
where $\mathcal{N}(x; \mu, \sigma)$ defines the probability at $x$ for a Gaussian probability distribution with mean $\mu$ and standard deviation $\sigma$.
We use these weights to randomly select 1200 realizations, forming the GWB-conditioned astrophysically realistic subset, used in \secref{sec:results_realpop}.
The rest of the results in \secref{sec:results_quickcw}--\secref{sec:results_confusion} use all 2000 samples: 400 from the low-$\snr$, $\classicuniform$ subset; 400 from the high-$\snr$, $\asa$ subset; and 1200 from the GWB-conditioned, $\asa$ subset.

\subsection{Pulsar Times of Arrival} \label{sec:methods_TOAs}
Given a $\holodeck$ population of binaries, the next step 
in analyzing CW detectability 
is generating the simulated TOAs that represent the response of our realistic PTA to this simulated GW universe. 
We use \texttt{pta-replicator} \citep{pta_replicator} for this, a tool that works with the standard PTA software \texttt{enterprise} \citep{enterprise} and \texttt{pint} \citep{pint2018, pint2024} to create and modify the TOAs of simulated pulsars. 
We set the observing times, TOA errors, and positions of the simulated pulsars to match those of the 67 pulsars used in the NG15 data analysis. 
Adopting the methods from \citet{pol+2021}, we use epoch-averaged TOAs and strip the simulated pulsars of their true residuals, as measured by \texttt{pint} \citep{pint2024}, before new Gaussian white noise, per-pulsar red noise, and the delays from the binary populations' GWs are injected.

The white noise injections correspond to the TOA measurement errors in the noise model of \citet{ng+23_timing}.
The red noise for each pulsar is constructed from a power-law spectrum \citep{ng+23_timing}
\begin{equation} \label{eq:toa_rednoise}
    P(f) = A_\tr{red}^2\bigg( \frac{f}{1 ~\tr{yr}^{-1}}  \bigg)^{\gamma_\tr{red}}
\end{equation}
with $A_\tr{red}$ and $\gamma_\tr{red}$ being the individual pulsar red noise values from \citet{ng+23_noise}.
Both noise components are sampled randomly from these distributions for each realization of each pulsar.

The GW emission is modeled in two parts following the approach of \citet{becsy+2022_realistic}: the loudest 100 single sources in each frequency bin and the background sum of all other sources. 
\citet{becsy+2022_realistic} and \citet{gardiner+2024} both showed that the outlier-subtracted spectrum is observationally isotropic for 10 or more outliers. 
Thus, this 100-outlier plus isotropic background approach is more than sufficient to fully resolve the characteristic strain spectrum. 

For the single-source component, we calculate the pulsars' response to each source's gravitational waveform, assuming circular orbits and randomly assigning positions, phases, inclinations, and polarization angles to each. 
For the background, we take the quadratic sum of sky- and polarization-averaged characteristic strains, again assuming circular orbits, of all remaining sources. 
Finally, the pulsar residuals are adjusted to include the effects of this background, based on the free spectrum amplitude at each frequency. 
With white noise, red noise, and GWs injected, the remaining TOAs represent a data set that matches the NG15 observing methods, but with time delays determined by the random (noise and binary population) realization.

\subsection{Signal-to-noise Ratios} \label{sec:methods_snrs}

The most direct way to assess CW detectability given both the injected signal, $\signal$, and the resulting TOA data including noise, $\data = \signal + n$, is to calculate a detection statistic like an $\snr$. 
Here, $\signal$ represents the timing residuals induced by a single circular CW source to 0th post-Newtonian order, calculated in \texttt{enterprise} following \citet[][Eqs.~1-7]{ng+23_individuals}, 
and $\data$ represents the pulsar timing residuals after modeling out all known delay components with \texttt{pint} so that only unmodeled-noise and gravitational-wave delays remain. 
The signal can also be written in terms of an overall  amplitude $\rho$ where $\signal = \rho \hat{\signal}$ and $\innerprod{\hat{\signal}}{\hat{\signal}}=1$.
The parenthetical notation refers to the noise-weighted inner product
    \begin{equation}
        ( \mathbf{a} | \mathbf{b} ) = \mathbf{a}^T \mathbf{C}^{-1} \mathbf{b}
    \end{equation}
for the noise covariance matrix $\mathbf{C}$, as calculated in \citet{becsy+2022_quickcw}.
We compare several methods for calculating $\snr$s to find the statistic that best corresponds to a full Bayesian analysis.

Our first $\snr$ is the \textit{optimal signal-to-noise ratio}, defined in \citet{jaranowski+2009} to be the $\snr$ that would occur if the emitted signal $\signal$ is optimally matched to the filter model $s$ such that $\signal = s$. 
Using the expectation value for $\rho$ in the presence of this signal, and variance $\sigma_{s=0}=\innerprod{\hat{\signal}}{\hat{\signal}}=1$ in the absence of the signal, \citet{dimatteo+2019} calculated
\begin{equation} \label{eq:sigsig}
    \sigsig \equiv \frac{\langle \rho_{\signal=s} \rangle }{\sigma_{s=0}} = \frac{\sqrt{\innerprod{s}{s}}}{\innerprod{\hat{\signal}}{\hat{\signal}}} = \sqrt{\innerprod{\signal}{\signal}}
\end{equation} 
In this work, it is true that $\signal=s$ because the injected CW signal follows the same assumptions and equations as our filter model (0th post-Newtonian order, circular orbit CWs). 
However, for individual realizations, the GWB noise and intrinsic pulsar red noise differ from the Gaussian ensemble-averaged variance of zero.
Thus, $\sigsig$ corresponds to an average detection statistic for the signal $\signal$ over many noise realizations, as opposed to any specific observation.
To improve on this, we identify $\snr$ calculations that include noise information via a $\data$ term.

Following the approach described in \citet{dimatteo+2019}, we consider the $\log$ of the ratio between the likelihood of observing the TOAs ($\data$) if the CW of form $\signal$ is present, to the likelihood of $\data$ if there is no CW signal present, 
as derived for Gaussian noise in \citet{ellis+2012}
    \begin{equation}
        \ln \Lambda = \ln \frac{\likelihood{\signal}{\data}}{\likelihood{0}{\data}} = \innerprod{\data}{\signal} - \frac{1}{2} \innerprod{\signal}{\signal}.
    \end{equation}
Maximizing this likelihood with respect to signal amplitude $\rho$ leads to the matched filter detection statistic \citep{dimatteo+2019},
\begin{equation}
    \rho_\tr{max} = \innerprod{\data}{\hat{\signal}} 
\end{equation}
and max-likelihood ratio
\begin{equation} \label{eq:matchedfilter_lnlike}
    \ln \Lambda_\tr{max} = \frac{1}{2} \rho_\tr{max}^2.
\end{equation}

This \textit{matched filter detection statistic} is our second $\snr$, as defined in \citet[][Eq.~7.47]{maggiore2007} to be
\begin{equation} \label{eq:sigdat}
    \sigdat \equiv \rho_\tr{max}=\frac{\innerprod{\signal}{\data}}{\sqrt{\innerprod{\signal}{\signal}}}.
\end{equation}
This detection statistic can be thought of as a general amplitude of the signal for the max-likelihood parameters.
Note that unlike $\sigsig$, $\sigdat$ incorporates information from the measured delays in addition to the injected signal.
For Gaussian noise, the average $\sigdat$ over many noise realizations would be equivalent to $\sigsig$.
We treat the true parameters of the injection as the max-likelihood parameters, noting that the likelihood may be maximized for slightly different parameters if those parameters better capture the noise.

Third, we can evaluate the maximum likelihood at our injected parameters, such that the expectation value for $\rho$ is precisely $\langle{\rho} \rangle = \rho_\tr{max}$, and apply Eq.\ \eqref{eq:matchedfilter_lnlike} to define the \textit{log-likelihood $\snr$}, often referred to as the reduced log-likelihood,
\begin{equation} \label{eq:lnlike}
    \snr_\Lambda \equiv \frac{\rho_\tr{max}}{\sigma_{s=0}} 
    = \frac{\sqrt{ 2\ln \Lambda_\tr{max}}}{\sigma_{s=0}} = \sqrt{2\innerprod{\data}{\signal} - \innerprod{\signal}{\signal}}. 
\end{equation}
This $\ln \Lambda_\tr{max}$ would match the $\mathcal{F}$-statistic likelihood ($\ln \Lambda$) used in the MCMC pipeline \citep[][Eq.~22]
{becsy+2022_quickcw} if the injected signal parameters are recovered accurately as the max-likelihood parameters. 
Previous works have identified $\ln \Lambda$ as the best statistic to decide whether or not a signal is present \citep{prix2007}.

We also test an approach that makes several approximations but can be calculated much faster by avoiding the use of any noise-weighted inner products. 
This fourth method uses an Earth-term-only ($\mathcal{F}_e$-statistic) $\snr$ for a uniform-duration and uniform-cadence PTA, following Eq.\ (35) in \citet{rosado+2015}. 
For each pulsar, we use the true sky location, white noise corresponding to the TOA measurement error, and red noise following the power law in \eqref{eq:toa_rednoise}.
Then, the duration for all of the pulsars is set to the maximum duration of the NG15 dataset, so that we include the lowest frequency bin, and a uniform observing cadence of 0.2 yr.
Ultimately, this method fails because it overestimates the detectability of sources in the lowest frequency bin, often overlooking detectable higher frequency sources in favor of the loudest source at the lowest frequency. 
This is a natural result of overestimating how many pulsars have data at the lowest frequencies, and is inconsistent with the more realistic finding that the best detected source is rarely in the lowest frequency bin, as shown later in \secref{sec:results_realpop} and \figref{fig:realistic_props}. 
Thus, using the full PTA noise-weighted inner products is necessary to make any prediction for a realistic CW search, so we exclude this method and focus our analyses on $\sigsig$, $\sigdat$, and $\lnlike$. 

To summarize these methods, $\sigsig$ describes the $\snr$ if the filter optimally matches the injected signal, averaged over many realizations of Gaussian noise; it is calculated using only $\signal$. 
$\sigdat$ is the amplitude of the signal when its likelihood for describing the data is maximized, and includes each specific realization's noise contributions to this amplitude from $\data$. 
$\lnlike$ is the $\snr$ of the max-likelihood scenario, assuming that to be at the injected parameters. 
$\lnlike$ and $\sigdat$ would agree, according to Eq.\ \eqref{eq:matchedfilter_lnlike}, if $\sigma_{s=0}$ were precisely 1. 
This would be the case when averaging over many white-noise-only scenarios or normalizing the white noise. 
In those cases, $\sigsig$ would also match the previous two because the noise would be orthogonal to $\signal$, such that $\innerprod{\signal}{\signal} = \innerprod{\signal}{\data}$. 
However, we include intrinsic pulsar red noise and GWB noise, making the three $\snr$ methods distinct.


\subsection{Bayesian MCMC Continuous Wave Searches}
\label{sec:methods_quickcw}

For our full CW searches, we adopt the methods used to place Bayesian limits on continuous waves in NG15 \citet{ng+23_individuals}, namely: \texttt{QuickCW} \citep{becsy+2022_quickcw}.
\texttt{QuickCW} builds on the MCMC approach of \texttt{enterprise}, fitting parameters in a common-uncorrelated-red-noise (CURN) + CW model, but 
\response{implements a custom likelihood calculation that precomputes the inner product for each set of shape parameters (those determining the morphology of the GW signal), resulting in highly efficient exploration of so-called projection parameters like initial phase, inclination, and polarization angle, which in turn, speeds up the entire analysis significantly.}
See \citet{becsy+2022_quickcw} for a full description of the model. 
The MCMC settings we select to optimize runtime and convergence are 100 million iterations, 8 chains with temperatures between 0 and 3, fixed pulsar red noise, a maximum CW frequency of 30 nHz (above which the white noise is too great for a realistic CW detection, as shown in \secref{sec:results_properties}), and a maximum chirp mass of $1\times 10^{11} \msol$. The remaining priors follow those of \citet{ng+23_individuals} and are given in \tabref{tab:priors}. 

\begin{deluxetable}{crr}
\label{tab:priors}
\tablecaption{Uniform $\qcw$ Priors}
\tablehead{
\colhead{Parameter} & \colhead{Minimum} & \colhead{Maximum}  
} 
\startdata
$\log_{10}(\fo /\tr{Hz})$ &	$-8.70$	& $-7.52$ \\
$\log_{10} (\hs)$               & $-18.0$ & $-11.0$ \\
$\log_{10} (\mc / \msol )$    & $7.0$ & $11.0$ \\
$\cos \theta$                   & $-1.0$ & $1.0$ \\
$\phi$                          & $0$ & $2\pi$ \\
$\cos \iota$                    & $-1.0$ & $1.0$ \\
$\log_{10} A_\tr{GWB}$  & $-18.0$ & $-11.0$ \\
$\gamma_\tr{GWB}$       & $0.0$ & $7.0$ \\
\enddata
\end{deluxetable}

The CW parameters whose recovery we will examine are the observed GW frequency $\fo$, the characteristic strain amplitude $\hs$, the observer-frame chirp mass $\mchirp$, the sky position with polar angle $\theta$ and azimuthal angle $\phi$ in celestial coordinates, the initial phase $\Phi_0$, the polarization angle $\psi$, and the inclination angle $\iota$. 
We also recover the GWB as a power-law CURN parameterized by the amplitude $A_\tr{GWB}$ at reference frequency $1~\yr^{-1}$ and spectral index $\gamma_\tr{GWB}$.

To evaluate whether a CW is found, we compare a model with a signal present, $\model_1$, and a noise-only model $\model_2$ (including the GWB). The Bayes factor between these models is given by the ratio of the probability density $p$ of the data given each model,
    \begin{equation} \label{eq:bf_def}
        \bfact = \frac{p(\data|\model_1)}{p(\data|\model_2)} 
    \end{equation}
The two models share all parameters $\commonpars$, except for the CW components. 
Following the Savage--Dickey approach, as outlined in \citet[][Chapter 6.3.5]{taylor2021}, we take a corner of parameter space in $\model_1$ where $\model_1$ matches the noise-only model $\model_2$. 
This is where the GW strain is the lowest allowed value, $\loghs=-18$, an amplitude several orders of magnitude below typical pulsar intrinsic noise, which sets the minimum on $\loghs$ priors. This null version of $\model_1$ replaces the noise-only $\model_2$, so that both share the same parameters, i.e.,
    \begin{equation} \label{eq:nullmodel}
        p(\data | \commonpars, \model_2) = p(\data|\loghs=-18, \commonpars, \model_1).
    \end{equation}
\citet{chatziioannou+2014} showed that the marginalized posterior can be written as 
\begin{equation} \label{eq:noise_posterior}
    p(\loghs=-18|\data, \model_1) = \frac{p(\data |\model_2) p(\loghs=-18)}{p(\data | \model_1)},
\end{equation}
which, combined with Eqs.\ \eqref{eq:nullmodel} and \eqref{eq:bf_def} gives the Bayes factor as the Savage--Dickey Density ratio,
\begin{equation} \label{eq:bf_savagedickey}
    \bfact = \frac{p(\loghs = -18)}{p(\loghs=-18 | \data; \model_1)}.
\end{equation}
The numerator and denominator represent the prior and posterior, respectively, of the lowest allowed amplitude.

In practice, the posteriors are given by a histogram of $N_\tr{tot}=500,000$ samples (100 million MCMC iterations with a thinning factor of 100 and burn-in of 50\%), uniformly spaced across the prior range $-18 \leq \loghs \leq -11$. 
We calculate the Bayes factor in \eqref{eq:bf_savagedickey} numerically as
\begin{equation} \label{eq:bf_numerical}
    \bfact = \frac{N_\tr{tot}}{N_\tr{1st\ bin} \cdot N_\tr{bins}}.
\end{equation}
Using 40 bins, the highest finite $\bfact$ we find is then $\frac{500,000}{(1)(40)}=12500$, which is sufficiently high to exceed any relevant Bayes factor threshold we might consider. 

We select the mode of the samples with 50\% burn-in to represent the recovered values $\commonpars_\tr{QCW}$, 
and describe the posterior width using the 68\% confidence interval (CI) width, $\ciwidth{\commonpars}$, calculated as the distance between the 16th and 84th percentile values, representing the 1 standard deviation region around the median. 

It is possible for the MCMC to converge on some set of parameters that do not truly represent a GWB and CW. 
For example the CW component can be fit to capture excess noise at low frequencies, either from the GWB or underestimated pulsar red noise. 
Thus, we must distinguish `true' detections from some narrow convergence on a model that describes the noise well, but is not representative of a real CW.

True detections tend to be distinguishable from false detections by their frequency recovery.
We set the conditions for a successful detection to be 
\begin{enumerate}
    \item[(1)] The $\fo$ posterior is well constrained: 
    \begin{equation} \label{eq:fo_width}
        \ciwidth{\fo}\leq \overline{\ciwidth{\fo}}
    \end{equation}
    for threshold width $\overline{\ciwidth{\fo}}=1.40~\tr{nHz}$, set by analysis of the data in \secref{sec:results_properties}. 
    \item[(2)] The injected $\fo$ falls within the recovered 99\% CI: 
    \begin{equation} \label{eq:fo_in_99}
        {f_{o,\tr{QCW}-}} \leq {\fo[\snr]} \leq {f_{o,\tr{QCW}+}}
    \end{equation} 
    or is accurate to within half a frequency bin: 
    \begin{equation} \label{eq:fo_delta}
        |\delta(\fo)| = 
        |f_{o,\tr{QCW}} - 
        f_{o, \snr} 
        | < 0.99~\tr{nHz}
    \end{equation}
\end{enumerate}   
We include the accuracy (second) condition because, occasionally, for very narrow posteriors, the injected $\fo$ can be very close to the recovered value without being within the 99\% CI. Thus, CWs are only required to pass Eq.\ \eqref{eq:fo_in_99} \textit{or} \eqref{eq:fo_delta} to be considered successfully detected. 


\subsection{$\snr$-Detection Fraction Mapping} \label{sec:methods_map}
With $\snr$s representing a quick detection statistic calculation for a set of simulated TOAs and $\bfact$ representing the result of a full MCMC search, we search for an efficient mapping from $\snr$ to $\bfact$ that is computationally cheaper, while preserving additional accuracy from the full Bayesian analysis. 
Naively, one might assume the log Bayes factors scale as $\snr^2$ as is the case when the posteriors are dominated by the likelihood.
However, most CWs relevant to this study have high enough $\snr$ that many parameters do not fill their prior volumes, causing the priors to play a greater role in the Bayes factor. 
Thus, the log Bayes factors follow an approximately linear dependence on $\snr$ \citep{littenberg+2016} with broad scatter. 
With full TOA modeling, the scatter in $\log_{10}\bfact$ versus $\snr$ often spans several orders of magnitude, making it unreasonable to fit a line directly (shown in Sec.~\ref{sec:results_map}). 
However, there is a clear trend between the $\snr$ and the detection fraction, where the detection fraction represents the number of samples at that $\snr$ with Bayes factors above a threshold value $\overline{\bfact}$.
Here, all samples are used, as opposed to just the true detections, because false detections also contribute to the number of detections identified by Bayesian analysis alone, without comparison to a known injection.

For an approximately linear dependence of $\log_{10} \bfact$ on $\snr$ ($\log_{10} \bfact \approx m \cdot \snr$) with Gaussian scatter $\sigma$, the log Bayes factors follow a normal distribution $\normal{m \cdot \snr}{\sigma}$. 
Thus, the probability of exceeding threshold $\log_{10} \bfthresh$ is found by integrating $\int_{\bfthresh}^\infty \normal{m \cdot \snr}{\sigma}$, such that the detection fraction (DF) follows a Gaussian cumulative distribution function ($\Phi$) of $\snr$,
\begin{equation}\label{eq:gcdf}
    \tr{DF}(\snr) = \Phi\left( \frac{m \cdot (\snr) - \log_{10} \overline{\bfact}}{\sigma} \right)
\end{equation}
With 2000 widely varying binary populations, binned in $\snr$ to have 25 points per bin, we use the \texttt{scipy.optimize.curve\_fit} \citep{scipy2020} function to perform a nonlinear least-squares fit for $m$ and $\sigma$.
We repeat this for each of the $\snr$ methods and compare the resulting reduced chi-squared values. 

\section{Results} \label{sec:results}

\subsection{Successful Detection Rates } \label{sec:results_quickcw}

\begin{figure}
    \centering\includegraphics[width=0.5\textwidth]{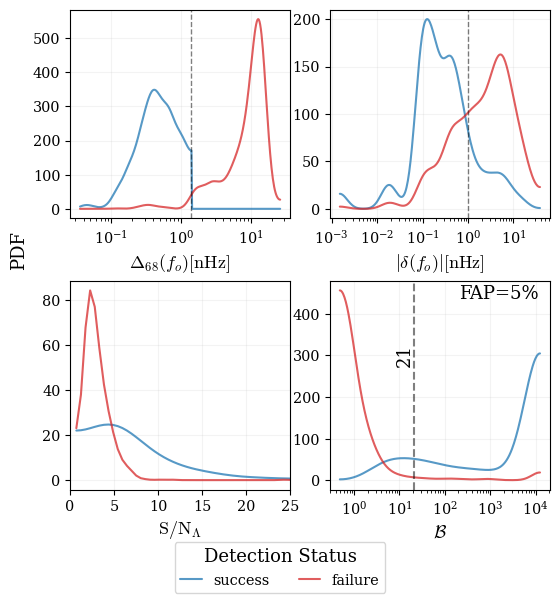}
    \caption{Distributions of successful (blue) and failed (red) $\qcw$ searches, as categorized by the criteria in Eqs.\ \eqref{eq:fo_width}--\eqref{eq:fo_delta}. The top left panel shows the precision of frequency recovery in terms of 68\% CI posterior width, and the top right panel shows the accuracy of frequency recovery as the distance between $\fo$ recovery and injection. Both have x-axes in nHz with the y-axes normalized to units of realizations per dex of nHz. 
    The bottom left panel shows the distributions of source $\lnlike$ (see Appendix \ref{sec:appendix_othersnrs} for $\sigsig$ and $\sigdat$ versions), with the y-axis as realizations. The bottom right panel shows the distributions of Bayes factor, with the y-axis as realizations per dex. The Bayes factor threshold required for a maximum false alarm probability of 5\% is marked by the dashed gray line. For the success distribution, the highest $\lnlike$ source meeting the detection criteria is used to determine the injected frequency and $\lnlike$. For the failed searches, the max-$\lnlike$ source is used.
    }
    \label{fig:detected_dist}
\end{figure}

Every $\qcw$ search is classified as a success or failure based on the frequency recovery conditions in \secref{sec:methods_quickcw}. 
For the accuracy criteria (Eqs.\ \eqref{eq:fo_in_99} and \eqref{eq:fo_delta}) we compare the recovered frequencies to the injected frequencies of the ten ``best" CWs, ranked by their $\lnlike$. 
By this classification, $\sim\!51\%$ of 2000 realizations contain a successfully detected CW source.

Figure \ref{fig:detected_dist} compares properties of the successful searches to the failed ones.
The distribution of $\fo$ posterior widths $\Delta_{68}\fo$ are shown for the successful samples in blue in the top left panel, and the failed samples are in red. 
The threshold $\overline{\Delta_{68} \fo}$ is marked by a dashed gray line, where the break between the two distributions is evident, supporting this threshold selection in the precision criterion, Eq.\ \eqref{eq:fo_width}. 

The top right panel shows $\absdeltafo$, the distance between injected and recovered  $\fo$. 
If multiple CWs meet the accuracy criteria within a realization that also passes the precision cut, then the max-$\lnlike$ source is used to calculate $\absdeltafo$.
The dashed vertical line shows the accuracy threshold required for a `successful' detection (Eq.\ \eqref{eq:fo_delta}), which is half of the width of a frequency bin. 
As described in \secref{sec:methods_quickcw}, narrowly recovered CWs can meet this accuracy condition even if their frequency is outside the 99 percentile region, when the 99\% CI spans $\lesssim 2~\tr{nHz}$ or is narrow and asymmetric. 
Even if there is a true detection, we sometimes misidentify which source this detection corresponds to. 
This can occur when the 68\% CI region is narrow (such that the precision criterion is met), but the 99\% CI region is broad and includes a higher-$\lnlike$ source than the one we more accurately recover. 
This explains the handful of successful detections with very high $|\delta (\fo)|$.
The occurrences of low $\absdeltafo$ in the failed distribution are a result of the max-$\lnlike$ source being coincidentally near the peak of the recovered histogram. 
This scenario is not uncommon given that the priors on $\fo$ have already been narrowed to the frequency regime where CW detections (and high $\snr$s) are feasible. 

The bottom left panel of \figref{fig:detected_dist} shows the $\lnlike$ distributions. 
Notably, there are some low-$\lnlike$ detections in our successful distribution, meaning that an $\lnlike<10$ does not rule out the possibility of detection by a full Bayesian analysis. 
However, none of the failed samples are above $\lnlike>10$, indicating that a very high $\lnlike$ is definitively predictive of a true detection. 
The equivalents of \figref{fig:detected_dist} are shown for $\sigsig$ and $\sigdat$ in Appendix \ref{sec:appendix_othersnrs}. 

The bottom right panel of \figref{fig:detected_dist} shows the Bayes factors of the successful and unsuccessful searches, conveying how often false detections by Bayesian analysis can occur.
If detections are identified according to their Bayes factor, the threshold for a false alarm probability (FAP) below 5\% is $\bfthresh_{05}=21$, marked by the dashed vertical line. 
Similar $\bfthresh_{05}$s for $\sigsig$ and $\sigdat$ are listed in \tabref{tab:gcdf}, along with the corresponding FAP and false dismissal probability (FDP) for this calibrated threshold $\bfthresh_{05}$, and fixed thresholds of 10, 100, and 1000. 

The maximum Bayes factor achievable by our methods is 12,500 (as explained in \secref{sec:methods_quickcw}), so the spike in the detected distribution represents all of the samples with $\bfact \geq 12500$.
This can occur for very strong detections, or false detections where the MCMC gets stuck on some false signal, a scenario discussed in \secref{sec:discussion}. 
The small spike appearing in the failed detection distribution is consistent with these high-$\bfact$ false detections being the most common false alarm scenario.
When using $\lnlike$ for CW selection, 1.7\% of the failed samples have Bayes factors above the upper limit, so an FAP below 1.7\% cannot be calculated. 
For a 2\% FAP, we falsely dismiss 44\% of the successful detections, and for a 5\% FAP, we falsely dismiss 21\% of the successful detections.

\subsection{$\snr$--Bayesian Detection Mapping} 
\label{sec:results_map}
\begin{figure}
    \centering
    \includegraphics[width=0.5\textwidth]{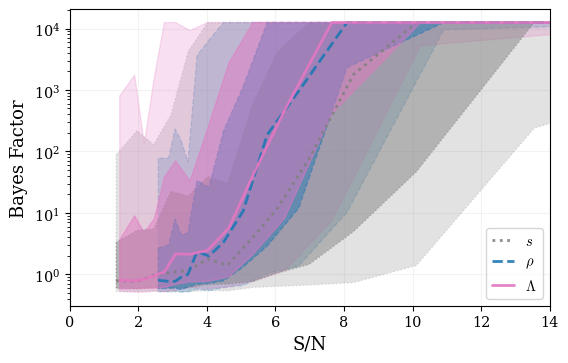}
    \includegraphics[width=0.5\textwidth]{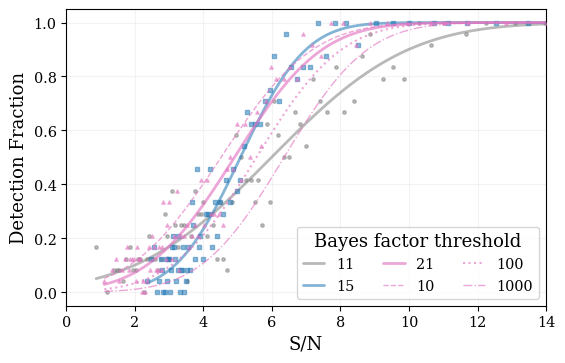}
    \caption{The top panel shows the median Bayes factor versus $\snr$ for $\sigsig$ (gray dotted line), $\sigdat$ (blue dashed line) and $\lnlike$ (pink solid line), with the 68\% CIs shaded. In the bottom panel, the scatter points represent the fraction of realizations in each $\snr$ bin meeting the 5\% FAP Bayes factor thresholds listed in the legend for $\sigsig$ (gray circles), $\sigdat$ (blue squares), and $\lnlike$ (pink triangles). The corresponding best-fit curves are shown as solid lines, with the Gaussian cumulative distribution function (GCDF) parameters listed in \tabref{tab:gcdf}. Additionally, the best fit curves for $\lnlike$ with $\bfthresh$ of 10, 100, and 1000 representing optimistic, realistic, and conservative thresholds are given in dashed, dotted, and dashed-dotted lines, respectively.
    }
    \label{fig:gcdf}
\end{figure}

\begin{deluxetable}{cccccccc} 
\label{tab:gcdf}
\tablecaption{$\snr$--Detection Fraction GCDF Fits}
\tablehead{
\colhead{Method} & \colhead{$\bfthresh$} & \colhead{$m$} & 
\colhead{$\sigma$} & \colhead{$\rchisq$} & \colhead{$\snr_{50}$} & \colhead{FAP} & \colhead{FDP}
} 
    \startdata
$\snr_s$	&	11	&	0.17	&	0.54	&	1.18	&	5.94	&	0.050	&	0.141	\\
	&	10	&	0.17	&	0.52	&	1.15	&	5.94	&	0.054	&	0.132	\\
	&	100	&	0.28	&	0.74	&	1.15	&	7.30	&	0.025	&	0.321	\\
	&	1000	&	0.36	&	1.00	&	1.01	&	8.27	&	0.009	&	0.431	\\
\hline  
$\snr_\rho$	&	15	&	0.23	&	0.35	&	1.34	&	5.12	&	0.050	&	0.176	\\
	&	10	&	0.20	&	0.31	&	1.19	&	4.84	&	0.058	&	0.131	\\
	&	100	&	0.35	&	0.52	&	1.18	&	5.81	&	0.029	&	0.321	\\
	&	1000	&	0.46	&	0.72	&	1.21	&	6.50	&	0.010	&	0.429	\\
\hline  
$\snr_\Lambda$	&	21	&	0.27	&	0.54	&	1.05	&	4.88	&	0.049	&	0.206	\\
	&	10	&	0.22	&	0.46	&	0.88	&	4.59	&	0.062	&	0.133	\\
	&	100	&	0.36	&	0.70	&	1.17	&	5.60	&	0.037	&	0.327	\\
	&	1000	&	0.47	&	0.90	&	1.72	&	6.33	&	0.021	&	0.438	\\
    \enddata
\tablecomments{Best-fit parameters for each fit to \eqref{eq:gcdf} by different $\snr$ methods and Bayes factor thresholds, $\bfthresh$}
\end{deluxetable}

Accurately mapping simulation information to realistic detectability is crucial to making the best estimates of CW occurrence rates over time, model constraints, and binary source predictions. 
To this end, we present a mapping relation between TOA-based $\snr$s and Bayesian detections and identify the best $\snr$ proxy.
While \figref{fig:detected_dist} shows the $\snr$ of the correctly identified CW, the $\snr$ mapping exclusively uses the highest $\snr$ in each realization to apply to scenarios where full posteriors are not calculated. 

In \figref{fig:gcdf}, the top panel shows the Bayes factor versus maximum $\sigsig$ (gray dotted), $\sigdat$ (blue dashed), and $\lnlike$ (pink solid) in each realization, with medians marked by dotted, dashed, and solid lines respectively, and the 68\% CI shaded.
\response{There are large discrepancies between S/N and Bayesian detectability attributed to limitations of the S/N calculations. Even $\sigdat$ and $\lnlike$ fail to perfectly capture individual noise realizations or wave interference effects because they use quadratically summed sky and polarization-averaged characteristic strains, not taking into account specific positions or phases of any source besides the signal source. These discrepancies are stochastic and cannot be attributed to any particular binary property or  GWB.}

In all cases, the scatter in Bayes factor versus $\snr$ is too great to fit a function reasonably, so instead we consider detection fraction versus $\snr$, shown in the bottom panel of \figref{fig:gcdf}. 
This detection fraction is defined as the fraction of realizations meeting each threshold $\bfact$.
\figref{fig:gcdf} contains the detection fraction in each $\snr$ bin using the $\overline{\bfact_{05}}$ thresholds for $\sigsig$ (gray circles), $\sigdat$ (blue squares), and $\lnlike$ (pink triangles). 
All CWs with $\sigsig \gtrsim 13$, $\sigdat \gtrsim 9$, or $\lnlike \gtrsim 9$ 
are detected. 
In the low- to intermediate-$\snr$ regime, we fit a Gaussian cumulative distribution function (GCDF) to these points, as described in $\secref{sec:methods_map}$, shown in the color of the corresponding $\snr$ method's scatter plot. 

In \tabref{tab:gcdf}, we present the best-fit parameters describing these curves: $m$ refers to the slope of a roughly linear $\log_{10} \bfact$ versus $\snr$ relation, $\sigma$ is the scatter for a normal distribution around this relation, as defined in Eq.\ \eqref{eq:gcdf}, and $\rchisq$ is the reduced chi-squared value. 
We use the $\bfthresh_{05}$ thresholds in addition to optimistic, realistic, and conservative thresholds of 10, 100, and 1000.
The $\snr$ required for a 50\% detection fraction is also listed, typically around $\sim\!5$--$6$, as well as the FAP and FDP for each Bayes factor threshold using the detected and null distributions in Figs. \ref{fig:detected_dist} and \ref{fig:detected_dist_appendix}.

Based on the reduced chi-squared statistic $\rchisq$ for each curve, $\lnlike$ is the best proxy in the 0.05 FAP ($\bfthresh_{05}=21$) and optimistic ($\bfthresh=10$) cases. 
Thus, we select $\lnlike$ as the fiducial $\snr$ for the rest of our analyses and show the $\lnlike$ best-fit curves for the remaining Bayes factor thresholds in \figref{fig:gcdf}. 
%
%
We apply the lowest $\rchisq$ relation, $\lnlike$ with $\bfthresh=10$, to make CW predictions for astrophysically realistic, GWB-conditioned populations in \secref{sec:results_realpop}.

\subsection{Binary Properties }\label{sec:results_properties}

\begin{figure}
    \centering
    \includegraphics[width=1.0\linewidth]{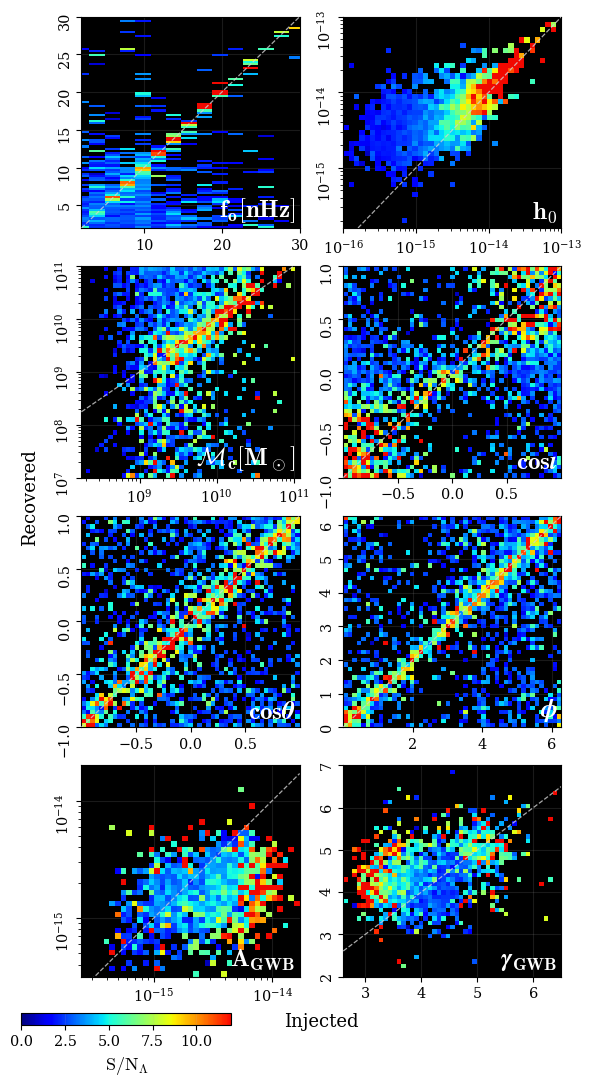}
    \caption{Each panel shows the recovery vs. injected value of one property, colored by $\lnlike$. The first row represents CW properties: observed GW frequency $\fo$ in nHz (left) and strain amplitude $\hs$ (right). The second row has the sources' chirp mass $\mc$ in $\msol$ (left) and inclination $\cos \iota$ (right). The third row has the source position in celestial coordinates: azimuthal position $\cos \theta$ (left) and polar angle $\phi$ in radians (right). The injected properties and $\lnlike$ are based on the highest-$\lnlike$ source detected in each realization or the max-$\lnlike$ source if none are detected. The last row contains GWB amplitude $\gwbamp$ and spectral index $\gwbgamma$, with the injection being based on a power law fit to the lowest 5 frequency bins of the injected $\hc$ spectrum, including all but the highest $\lnlike$ source.
    }
    \label{fig:props_rec_vs_inj}
\end{figure}

\begin{figure}
    \centering
    \includegraphics[width=1.0\linewidth]{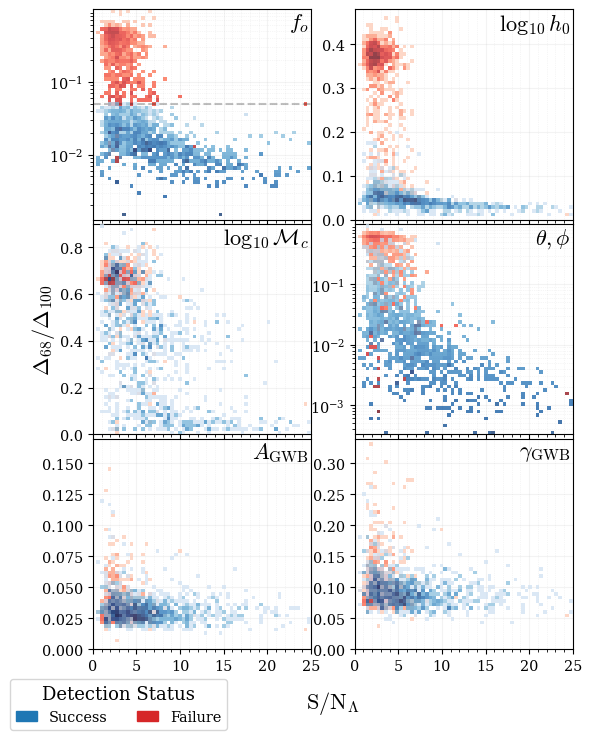}
    \caption{The y-axes show 68\% CI posterior widths as a fraction of total prior width $\Delta_{68}/\Delta_{100}$, and the x axis is the $\lnlike$ of the detected, or highest-$\lnlike$ nondetected CW in each realization.  This precision versus $\lnlike$ is shown for  CW properties $\fo$ (top left), $\loghs$ (top right), and $\mc$ (middle left); localization area (middle right), and GWB amplitude at $f_\tr{ref}=1/\yr$ (bottom left) and spectral index (bottom right). The blue 2D histograms represent successful detections while the failed searches are in red, matching the groupings in \figref{fig:detected_dist}.
    The dashed line in the top left panel shows the threshold posterior width ($\overline{\Delta_{68}/\Delta_{100}(\fo)}=0.05$ or $\overline{\Delta_{68}(\fo)}=-1.4~\tr{nHz}$) used in \secref{sec:results_quickcw} to distinguishing successful detections by Eq.\ \eqref{eq:fo_width}.
    Similar plots with groupings determined by $\sigsig$ and $\sigdat$ sorted CWs are in \figref{fig:props_ci_vs_snr_appendix}. 
    }
    \label{fig:props_ci_vs_snr}
\end{figure}

\begin{figure}
    \centering
    \includegraphics[width=\linewidth]{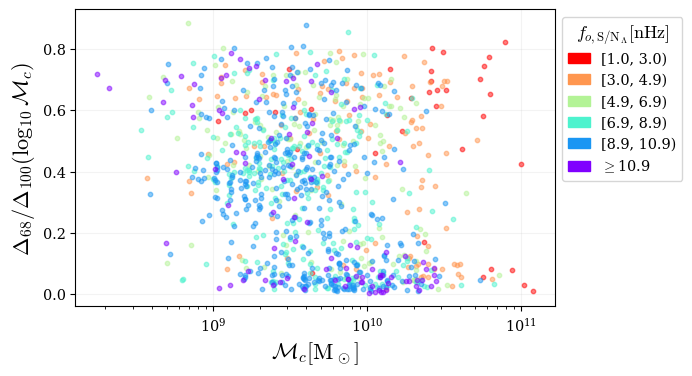}
    \caption{The y-axis shows the chirp mass recovery precision for successful detections as the normalized 68\% CI posterior width of $\logmc$, matching the blue data in the middle left panel of \figref{fig:props_ci_vs_snr}. The mass precision is plotted against injected mass and colored by injected frequency to demonstrate how frequency evolution (faster for higher masses and higher frequencies) impacts chirp mass measurement.
    }
    \label{fig:mc68_vs_mcinj}
\end{figure}

Once a CW is detected, the next step will be to identify the binary source properties, potentially aiding in the search for an EM counterpart and multimessenger analysis.  
We examine how well the binary properties can be recovered first by comparing the posterior modes of frequency $\fo$, strain amplitude $\hs$, chirp mass $\mchirp$, sky coordinates $\cos\theta$ and $\phi$, and inclination $\iota$, to the corresponding injected values in \figref{fig:props_rec_vs_inj}. 
For realizations with multiple CWs meeting the detection criteria, the highest-$\lnlike$ one is selected to determine the injected properties. 
Similarly, when none are detected, we use the highest-$\lnlike$ source for analyses of the undetected sources.
The bottom row depicts the amplitude $\gwbamp$ and spectral index $\gwbgamma$ recoveries.
The injection values on the x-axes are based on a power-law fit to the first five frequency bins of the injected characteristic strain spectrum of all but the highest-$\lnlike$ source.

The points in \figref{fig:props_rec_vs_inj} are colored by $\lnlike$, showing that almost all $\lnlike$ $\gtrsim 7$ sources have very well-recovered $\fo$, $\hs$, and sky position.
There are few intermediate $\lnlike$ samples at $\fo \gtrsim 15 \tr{nHz}$ because each realization's highest- $\lnlike$ CW is rarely at high frequencies, where loud CWs are less common, but when they do occur, they have less red noise to compete with.
Low $\lnlike$ samples do occur at these high frequencies because if the realization doesn't have any especially loud CWs, then a high frequency source could have the max $\lnlike$, but it is unlikely to be detected. 
These represent universes with few high-mass SMBH outliers, where a CW detection is the least likely. 
Nondetection posteriors tend toward lower frequencies, where red noise dominates. 
In particular, when the GWB is modeled without spatial correlations (as with $\qcw$), some of this noise can be picked up as a false low-frequency CW detection. 
This scenario could explain the four false detections in the lowest frequency bin ($0.9$--$2.9~\tr{nHz}$) out of the 48 failed searches with Bayes factors above the FAP-calibrated threshold $\bfthresh=21$.
These are usually distinguishable from true detections by their wider posterior distributions for non-frequency parameters.

\figref{fig:props_ci_vs_snr} shows the width of the posteriors' 68\% CIs versus $\lnlike$ for $\fo$ (top left), $\hs$ (top right), $\mc$ (middle left), $\gwbamp$ (bottom left), and $\gwbgamma$ (bottom right) as a fraction of the prior width, $\Delta_{68}/\Delta_{100}$. 
The confidence on $\theta$ and $\phi$ are conveyed by the 68\% confidence localization area as a fraction of the whole sky (middle right). All are given as 2d histograms of the successfully detected distribution in blue, and the failed searches in red, corresponding to the same groupings as in \figref{fig:detected_dist}.
\figref{fig:props_ci_vs_snr} shows that almost all the high-$\lnlike$ samples meet the successful detection criteria (in blue) with narrow posteriors of $\ciwidth{\fo} \lesssim 0.5 \tr{nHz}$, $\ciwidth{\loghs}\lesssim 0.4~\tr{dex}$, and $\ciwidth{\theta,\phi} \lesssim 1000~\tr{deg}^2$).
There is a distinguishable grouping of all narrowly-recovered, high $\lnlike$ samples having $\ciwidth{\fo}$ less than 5\% the prior width, marked by the dashed gray line in the top left panel of \figref{fig:props_ci_vs_snr}. 
Thus, we use this cutoff, corresponding to $\ciwidth{\fo}=1.4\tr{nHz}$ as the threshold for the successful detection criteria in Eq.\ \eqref{eq:fo_width}.

At lower $\lnlike$ the CW parameters can be similarly well recovered (in both accuracy and posterior width) when detections are successful, but there are a large number of undetected samples with $\Delta_{68}/\Delta_{100}$ clustering around 0.5 for $\fo$, $0.38$ for $\loghs$, $0.75$ for $\mchirp$, and $0.65$ for sky position.
The injected inclination, randomly selected for every binary, tends toward a face-on orientation ($\cos \iota = \pm 1$), since that maximizes GW amplitude. The recoveries can be accurate, but they sample the parameter space more uniformly, often being more edge-on ($\cos \iota = 0$) than the injection. 
%
When $\hs$ is inaccurately recovered, it tends to be overestimated, especially for weaker injections. 
At higher $\lnlike$s, this excess strain can be absorbed by recovering more edge-on inclination than the injection.
For nondetections, the overestimation can just be a result of posterior modes tending toward the middle of the priors, $\loghs=-14.5$. 

The GWB recovery is generally similar for detected and nondetected populations, but tends to be better recovered for very high $\snr$, where the GWB and CW strains are typically both louder.
Notably, in \figref{fig:props_rec_vs_inj}, the highest amplitude GWB ($\gwbamp \gtrsim 4 \times 10^{-15}$) injections have low $\lnlike \lesssim 2.5$, where the GWB noise competes with even the loudest CWs. 

The chirp mass is also often well recovered for the high-$\lnlike$ samples, but can be drastically underestimated even in some very high $\snr$ scenarios. 
This typically occurs when the CW frequency is low. 
\figref{fig:mc68_vs_mcinj} show $\Delta_{68}/\Delta_{100}$ of $\logmc$ versus injected $\mc$, colored by injected $\fo$. 
There, the only cases where very high mass ($\gtrsim 3\times 10^{10}~\msol$) CWs lack narrow mass recoveries are when they are in the first or second frequency bin ($1.0$--$4.9~\tr{nHz}$).
At lower masses, most low frequency CWs have broad mass posteriors.
This is because the chirp mass is encoded in the rate of frequency evolution, which increases at higher frequencies and for higher masses. 
Thus, although high mass, high frequency sources are rare, they are the most likely to have successful mass measurements. 
Below $\sim2\times10^{10}~\msol$, no CWs in the first frequency bin have narrowly-constrained masses, and only a few in the second frequency bin do. 
The second row, left panel of \figref{fig:props_rec_vs_inj} shows that in the poor recovery scenarios, chirp masses tend to be underestimated. 
A trans-dimensional model could address this by allowing for the inclusion or exclusion of the frequency evolution parameter, in effect removing the chirp mass parameter when it is not constrained.

\subsection{Multiple Source Confusion} \label{sec:results_confusion}

\begin{figure}
    \centering
    \includegraphics[width=0.5\textwidth]{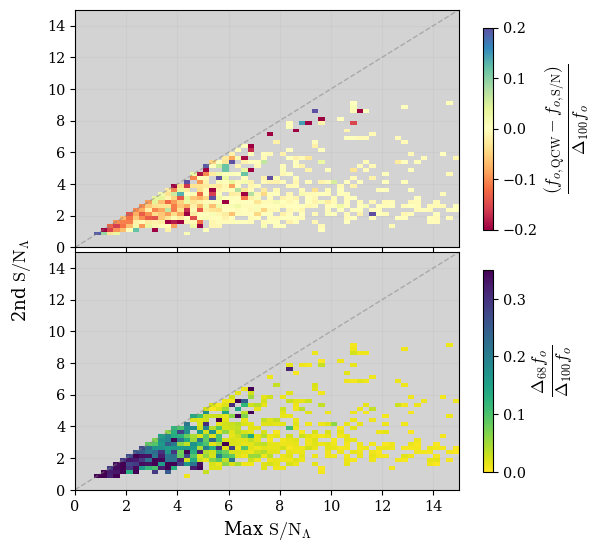}
    \caption{The second highest $\lnlike$ versus maximum $\lnlike$, colored by accuracy (top panel) and precision (bottom panel). The top panel shows that when there are multiple high-$\lnlike$ sources present, simulations can incorrectly predict which source will be detected, indicated by large gaps between recovered $f_{o,\tr{QCW}}$ and injected $f_{o,\lnlike}$ (marked by darker colored points). The bottom panel shows that high $\lnlike$ realizations can also have wide posteriors when the second highest $\lnlike$ is competitive with the first. }
    \label{fig:source_confusion}
\end{figure}

One particular scenario that can prevent accurate CW source recovery by any GWB+CW or CURN+CW model is source confusion between multiple loud CWs.
We recover a CW ranked second or higher in $\lnlike$ about 10\% of the time, preventing agreement with the max-$\lnlike$ source. 
This is confirmed in the top panel of \figref{fig:source_confusion}, where we show the second highest $\lnlike$ versus the max $\lnlike$ of each realization, colored by the difference between recovered $\fo$ and max-$\lnlike$ injected $\fo$ 
($\delta (\fo) \equiv 
\fo[\tr{QCW}] - \fo[\lnlike]$) 
as a fraction of prior width.
It is evident that above $\lnlike \gtrsim 7$, inaccurate $\fo$ recoveries only occur when the second highest $\lnlike$ is also high.

Not only are competitive sources common, but we demonstrate that this can prevent a narrow recovery of either source in the bottom panel of \figref{fig:source_confusion}. 
There, the points of second versus maximum $\lnlike$ are colored by the 68\% CI width. Above $\lnlike\gtrsim6$, poor recoveries marked by wide posteriors are rare. 
However, several of these cases do occur, and they are almost always coincident with a competitive second source. 
In \secref{sec:discussion} we discuss these effects in more detail for specific competitive source scenarios,  and suggest ways to solve this with multi-CW search models \citep[e.g][]{becys+cornish2020, taylor+2020}.

\subsection{\response{GWB-conditioned} Populations}\label{sec:results_realpop}

\begin{figure}
    \centering    \includegraphics[width=0.5\textwidth]{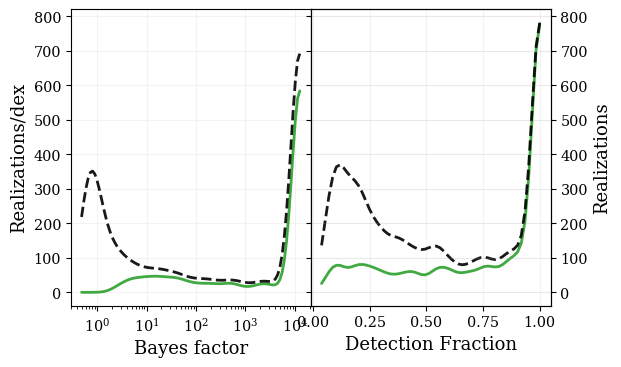}
    \caption{Bayes factor (left) and detection fraction (right) distributions in GWB-conditioned populations constructed from astrophysically realistic semi-analytic models (SAMs). The dashed black line represents all populations in this conditioned subset, while the solid green line only includes the successful detections, meeting the Eqs.\ \eqref{eq:fo_width}--\eqref{eq:fo_delta} criteria. 
    The detection fraction is calculated using the minimum-$\rchisq$ mapping from $\lnlike$ to fraction with $\bfact > 10$. 
    }
    \label{fig:realistic_bfs_dfs}
\end{figure}

\begin{figure}
    \centering
\includegraphics[width=0.5\textwidth]{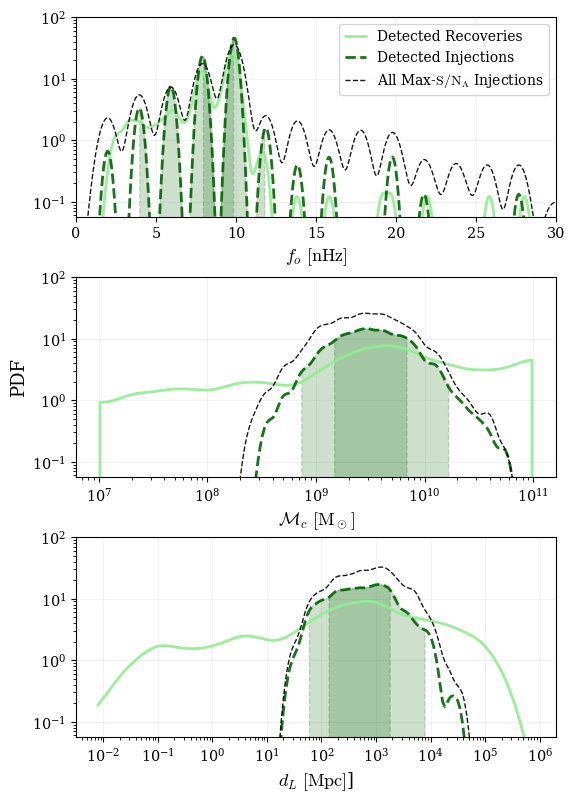}
    \caption{Binary properties found in the conditioned subset of populations, including observed GW frequency $\fo$ (top), chirp mass $\mc$ (middle) and luminosity distance $\dl$ (bottom). The dashed black line represents the property of the max-$\lnlike$ CW in every realistic population. The green lines represent only the successfully detected CWs, based on the frequency recovery criteria (Eqs.\ \eqref{eq:fo_width}--\eqref{eq:fo_delta}), with the dashed line being the injected value and the solid line being the recovered value. The shaded green regions are the 68\% and 95\% confidence intervals on detected injections, representing the true binary properties of possible CW detections.
    }
    \label{fig:realistic_props}
\end{figure}

Finally, we apply our analyses to a subset of binary populations consistent with the NG15 GWB, drawn from the astrophysically motivated parameter space as described in \secref{sec:methods_pops} and \tabref{tab:asa_pspace}.
\response{While there is ongoing tension between favored astrophysical models and NG15 GWB measurement, we consider this subset of samples to be the most realistic because it incorporates both the GWB and astrophysical priors.}
Given that most models predict GWB amplitudes below the NG15 measurement, these samples are skewed to the lower end of the NG15 error range, with $A_\tr{GWB}$ at $f_\tr{ref}=\frac{1}{\yr}$ spanning $5.0\times 10^{-16}$ to $8.4\times10^{-15}$, with the middle 95\% between $9.6\times 10^{-16}$ and $5.5\times 10^{-15}$. 

Fifty-three percent of these samples are successfully recovered by the $\fo$ recovery criteria (Eqs.\ \eqref{eq:fo_width}--\eqref{eq:fo_delta}), with the Bayes factor distribution shown in the left panel of \figref{fig:realistic_bfs_dfs}. 
Assuming the observed GWB amplitude on which we conditioned this subset is correct, there is some tension between this high detection fraction and the current lack of evidence for a CW in \citet{ng+23_individuals}. 

The dashed black line represents the entire conditioned subset while the solid green line only includes the successfully recovered samples. 
The distribution peaks at $\bfact \sim 1$ and the upper limit of $\bfact \geq 12500$, while Bayes factors between 10 and 1000 rarely occur. 
Thus, these populations tend to have a clear strong detection or no detection. 
$\sim\! 50\%$ have Bayes factors below 10, consistent with the current lack of high values in \citet{ng+23_individuals}. 
These detection rates \response{may be sensitive} to GWB amplitude and shape.

The right panel in \figref{fig:realistic_bfs_dfs} shows the histogram of each group's detection fraction, based on the minimized $\rchisq$ mapping from $\lnlike$ to detection fraction exceeding $\bfthresh = 10$. 
This demonstrates how we can recover a distribution similar to the Bayesian analysis using the maximum $\lnlike$'s and the GCDF mapping. 

Figure~\ref{fig:realistic_props} contains the distributions of $\fo$ (top panel), $\mc$ (middle panel), and $\dl$ (bottom panel) across this conditioned subset of samples. 
The dashed green line represents the true properties of detectable CWs, i.e. the max-$\lnlike$ source meeting the Eqs.\ \eqref{eq:fo_width}--\eqref{eq:fo_delta} frequency recovery criteria in each successful realization. 
68\% (and 95\%) of these realizations have 
GW frequencies 
between \response{$7.9$ and $9.9~\tr{nHz}$ ($4.0$ and $11.9~\tr{nHz}$), 
chirp masses 
between $1.5~\times~10^{9}$ and $7.2~\times~10^{9}~\msol$ ($7.1~\times~10^{8}$ and $1.7~\times~10^{10}~\msol$),
and luminosity distances 
between $138$ and $1941~\tr{Mpc}$ ($56$ and $8053~\tr{Mpc}$).}
The injected $\holodeck$ populations are binned in frequency based on the last NANOGrav dataset duration of 16.03 yr. 
Since all sources in a given bin are assigned the midpoint $\fo$ value, these bins are reflected in the injected distribution, where most realizations are in the second through sixth bins, above which white noise dominates the signal. 
However, occasional detections up to $28~\tr{nHz}$ are possible.

The solid green line corresponds to the $\qcw$--mode of each property. 
The difficulty in recovering $\mc$ for low frequencies is reflected by a very wide $\mc$ distribution,
spanning the allowed range of $10^7~\msol - 10^{11}\msol$. 
Since the injections are rarely more massive than $4\times10^{10}\msol$, the highest mass measurements tend to be overestimated, corresponding to the overestimated luminosity distances $\gtrsim 4\times 10^4$~Mpc in the bottom panel.
Meanwhile, the underestimated chirp masses produce unrealistically close distances, reaching $\dl \lesssim 25~\tr{Mpc}$, many of which can be ruled out by galaxy catalogs \citep{petrov+2024}. 

The dashed black line represents the max-$\lnlike$ injections of all realizations in the conditioned subset of populations, detected or not. 
High $\lnlike$ sources are often successfully detected at low-frequencies. Meanwhile, when the max-$\lnlike$ source is at high frequencies, it remains difficult to detect. 
This suggests that realistically detectable sources are at low frequencies.

\begin{figure}
    \centering
    \includegraphics[width=0.5\textwidth]{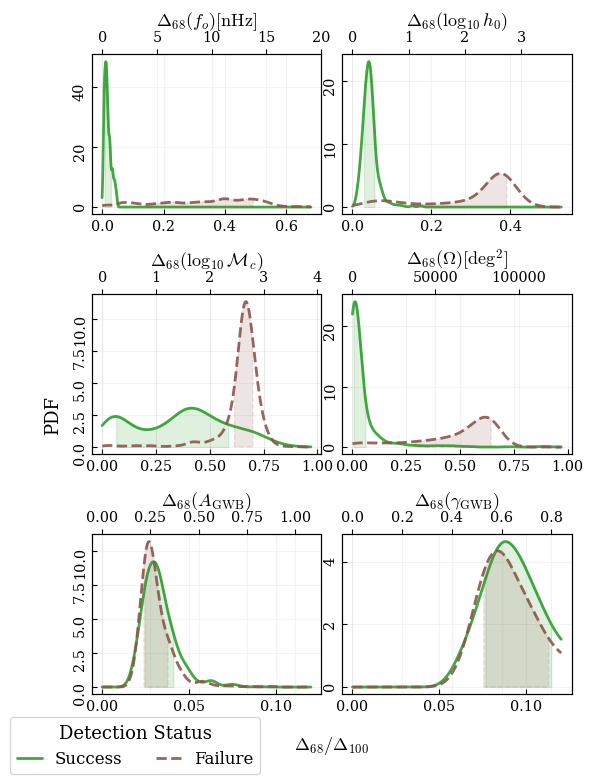}
    \caption{
    Each panel shows a histogram of the 68th percentile posterior width $\Delta_{68}$ on the top axis and the corresponding normalized width $\Delta_{68}/\Delta_{100}$ for the parameter in the same panel of \figref{fig:props_ci_vs_snr}. The histogram distributions are split into all successful detections (solid green) and failed searches (dashed brown) in the realistic GWB-conditioned subset, with each 68\% CI shaded in the same color. 
    }
    \label{fig:realistic_cis}
    
\end{figure}

Figure~\ref{fig:realistic_cis} shows the posterior widths corresponding to the same parameters as in each panel of \figref{fig:props_ci_vs_snr}, for successful detections (solid green) and failed searches (dashed brown), with the 68\% CIs shaded.
The detected realizations all have narrow recoveries on $\fo$, $\hs$, and sky position.
The median 68\% CI posterior width of $\logmc$ for successful detections is $1.53~\tr{dex}$, with 68\% between 0.26 and 2.35 dex and 95\% between 0.07 and 3.01 dex.
While the $\mc$ recoveries of detections vary, there is a typical width of failed detections around $\Delta_{68}/\Delta_{100}(\log_{10}\mc)$ of $\sim\! 68\%$, as expected for a uniform probability distribution matching the priors.
The same is true for localization. 
The 68\% CI width of failed searches' $\loghs$ is lower, indicating narrower posteriors than priors, favoring low $\hs$.  
The GWB recovery is similar, with CI's on both $\gwbamp$ and $\gwbgamma$ near 2-5\% of the prior, for all samples regardless of whether a CW is detected. 
This is unsurprising given that these samples are all drawn from a distribution around the same GWB amplitude and slope. 
The consistency between these distributions suggests that recovery of a GWB consistent with NG15 is successful when we include a CW in the search model, regardless of whether a CW is also present. 
This offers support for using a CW+GWB search model, which solves the problem of poor GWB recovery when a GWB-only search is performed with a CW present in the data \citep{becsy+2023_howtodetect}. 

\begin{figure}
    \centering
    \includegraphics[width=\linewidth]{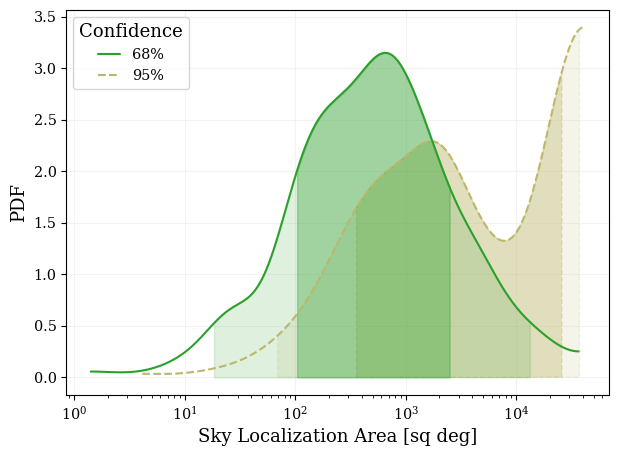}
    \caption{Localization of successfully detected CWs in the GWB-conditioned subset of populations. The solid green curve is a zoom-in on the detected distribution of $\theta, \phi$ (middle right) panel of \figref{fig:realistic_cis}, representing 68\% confidence localizations areas converted to sq deg. The dashed yellow curve represents the 95\% confidence localization areas of these samples. The shaded regions correspond to the 68\% and 95\% CI on each distribution of areas, among all realizations.  
    }
    \label{fig:realistic_localization}
\end{figure}

\figref{fig:realistic_localization} zooms in on the localization areas of the successful detections in square degrees, with 68\% confidence localization areas in green and 95\% confidence localization areas in yellow. 
Their medians are 537 $\tr{deg}^2$ and 2196 $\tr{deg}^2$, respectively.
The shaded regions show the 68\% and 95\% CI ranges among all realizations of the 68\% (and 95\%) confidence areas, with 68\% of realizations spanning $103-2,519~\tr{deg}^2$ ($356-25,961~\tr{deg}^2$) and 95\% of realizations between $18-13,274~\tr{deg}^2$ ($68-37,910~\tr{deg}^2$).

\section{Discussion} \label{sec:discussion}

\subsection{Limitations of CW Searches} \label{sec:disc_limitations}
We conduct Bayesian CW searches on $\sim$2000 simulated TOA datasets generated by calculating the PTA response to full binary populations. 
By modeling these discrete populations, 
we can identify limitations to current methods that have been less apparent in studies using a power-law approximation for their GWB injections. 
For example, \secref{sec:results_confusion} shows that a second high $\snr$ source impacts the recovery of the maximum $\snr$ source.
A clear understanding of how competitive sources might appear in CW searches requires more detailed investigation, but we offer some observed scenarios and plausible explanations to guide future work.

Several realizations narrowly recover parameters disparate from any of the high $\snr$ injected sources. 
This could correspond to a beat frequency between multiple sources, most likely when the recovered frequency is higher than those injected. 
Other realizations miss a high $\snr$ CW, but converge on a frequency hosting multiple intermediate-to-low-$\snr$ sources. 
If the phases align, the multiple sources could produce an amplified signal through constructive interference. 
Finally, destructive interference could prevent a narrow recovery of any source, as may be the case in the wide $\ciwidth{\fo}$ cases of \figref{fig:source_confusion}. 

While a GWB + two CWs model might be the best fit to some of these competitive source scenarios, there are plenty of other cases where only one CW is detectable. 
Thus, we recommend a trans-dimensional model approach that compares single-CW and multi-CW models. 
\citet{becys+cornish2020} presented a Bayesian inference method that uses reversible jump MCMC to jump between models with different numbers of CWs and \citet{gunderson+cornish2025} discussed how the reversible jump MCMC could be used with their Fourier basis method to speed up trans-dimensional searches.  

Multiple CWs could also be distinguished spatially by power-based anisotropy methods like the point-source approach in \citet{taylor+2020} which determines the number of point sources over an isotropic background that best fits the data, according to Bayesian analysis. 
If the CWs are at different frequencies, they could also be distinguished by the per-frequency optimal statistic. 
\citet{gersbach+2025} showed that this method can identify a single-frequency spike in amplitude corresponding to an injected CW, and in principle, the same could be done to identify multiple CWs if they are at different frequencies. 

We also gain insight into false detection scenarios related to confusion between the GWB and CW components of the model.
Given that $\qcw$ uses a CURN+CW model, it is possible that the lack of spatial correlations in the GWB component makes it easier for the CW component to capture some of the GWB as a false detection. 
This is most obvious in certain high $\bfact$, low $\snr$ cases where GWB strain in the lowest frequency bin is misidentified as a CW and the remaining GWB is recovered with a nonphysical shape, e.g.~an unreasonably flat or positively sloped characteristic strain spectrum. 
In contrast, the multiple CW confusion cases still recover a reasonable background. 
Thus, the recovered frequency and GWB slope can distinguish these two scenarios.
Poor recoveries of $\theta$, $\phi$, and $\iota$ are also associated with false detections, but they are not definitive indicators because intermediate $\snr$ CWs can be detected but poorly localized \citep[e.g.][]{petrov+2024}. 

The misidentification of GWB strain as a CW could be avoided by including HD correlations in the GWB component of the model \response{\citep{epta+23_gwb, ferranti+2024}}. 
However, computing a CW+HD likelihood within the MCMC \response{for many realizations} is prohibitively slow. 
\citet{ng+23_individuals} address this by using the likelihood re-weighting technique in \citet{hourihane+2023} to account for spatial correlations in post-processing. 
While a low post-processed likelihood can accurately discount false detections, a true CW may be missed due to convergence on this false CW+CURN model.
This reweighting technique is more likely to fail for high $\snr$ cases where the posteriors are already narrow.

\response{
It is also possible that some false detections are capturing inaccurately modeled pulsar red noise. 
\citet{hazboun+2020} and \citet{dimarco+2025} showed how pulsar red noise modeling can have significant impacts on GWB recovery. 
Investigating whether the pulsar red noise models also impact CW recovery may help limit these false detections.
}

\subsection{Comparison to Previous NG15 CW Upper Limits} \label{sec:disc_compare}

This work is uniquely comparable to current observations in that the binary populations are based on astrophysical models consistent with the astronomical literature/observations; the TOAs are matched to the pulsar positions, measurement noise, and observing times of NG15; and samples are conditioned on their agreement with the NG15 GWB measurement. 
Thus, we can directly compare our results to the CW upper limits in \citet{ng+23_individuals} \response{and closely compare to the European pulsar timing array CW search \citep{epta+23_individuals}.}

Although two candidate signals were investigated, \citet{ng+23_individuals} ultimately found no compelling CW evidence. 
This contrasts with our finding that 53\% of populations conditioned on the NG15 GWB contain successfully detected CWs and 50\% have $\bfact \geq 10$. 
It remains plausible that our Universe is one of the 47\% of cases without a CW detection. 

If CW nondetection continues while sensitivity improves, this tension could build, forcing us to consider alternate GWB models. 
A simple explanation could be that the current GWB amplitude is vastly overestimated. 
In that case, the GWB-conditioning in our sample selection favors samples with unrealistically high characteristic strain spectra, correlated with higher single-source strain amplitudes. 
The combined International Pulsar Timing Array Data Release 3 will provide some insight into the reliability of the current amplitude measurement. 
The other possibility is that the GWB amplitude contains non-SMBHB contributions, whether excess noise or non-binary sources of GWs \citep{ng+23_newphysics}. 
In a follow-up paper, we will apply these methods to extended PTA durations, and place tighter constraints on the time at which CW nondetection would become inconsistent with a GWB produced primarily by SMBHBs.

Although the NG15 PTA is most sensitive to 6~nHz  (third frequency bin) CWs \citep{ng+23_individuals}, 
when incorporating binary population modeling, 82\% of the detected realistic CWs (\figref{fig:realistic_props}) are between $7$ and $11~\tr{nHz}$ (fourth or fifth frequency bin).
The 95\% CI spans $4$--$12~\tr{nHz}$, which does include the frequency where \citet{ng+23_individuals} identified weak evidence ($\bfact\sim3$) for a CW at $\sim\!\!4~\tr{nHz}$. However, that CW became disfavored when they incorporated HD correlations into the likelihood.
\response{
Similarly, \citet{epta+23_individuals} found a strong favoring of the CW + noise over the noise-only model at 4.6~nHz source ($\bfact \sim 4000$), but this Bayes factor decreased to $\sim\!\!10$ when including CURN, and $\sim\!\!1$ when including HD correlations.
}
\citet{ng+23_individuals} also found evidence ($\bfact \sim 300$) for a $\sim\!\!170~\tr{nHz}$ candidate coinciding with the binary period of pulsar J1713+0747, which weakened by an order of magnitude when this pulsar was excluded.
The rarity of high-frequency detections in our results (\figref{fig:realistic_props}) favors the conclusion that this excess likelihood at $\sim\! \!170$~nHz is not a real CW.


\section{Conclusions} \label{sec:conclusion}
This work establishes methods for testing the detectability of Continuous Waves (CWs) in simulated binary populations by Bayesian analysis. 
We apply these methods to address how well CWs can be detected, how accurately and precisely their sources' binary properties can be measured, and what those properties are most likely to be. 
Additionally, we develop the tools to address what constraints CW upper limits can place on astrophysical models and what the CW detection prospects are over time. Our key findings are as follows:

\begin{enumerate}
    \item We find that 53\% of binary populations that are conditioned on the NG15 GWB measurements contain a successful CW detection.  This implies that if the measured GWB properties are correct and CWs remain undetected over the next few years, it may indicate tension with the SMBH binary model.

    \item Among the realistic, GWB-conditioned populations, 95\% have GW frequencies between $4$ and $12~\tr{nHz}$, chirp masses between \response{$7.1~\times~10^{8}~\msol$} and $1.7~\times~10^{10}~\msol$, and luminosity distances between $56~\tr{Mpc}$ and $8.1~\tr{Gpc}$.  
    Additionally, 95\% of these realizations have 68\% (and 95\% ) confidence sky localization areas between $18$ and $13,274~\tr{deg}^2$ ($68$ and $37,910~\tr{deg}^2$) and 68\% CI chirp mass posterior widths between $0.07$ and $3.01$ dex.
    Chirp masses are often poorly recovered, with only 29\% of these successful detections identifying the chirp mass accurately to within 1 dex with a 68\% posterior width also narrower than 1 dex.
    
    \item 
    We develop a simple mapping to rapidly calculate a Bayesian-informed detection statistic. 
    One can calculate a reduced log-likelihood ratio between signal and noise models, calculated at the injected parameters, and convert this to a detection fraction using the Gaussian cumulative distribution function (Eq.\ \eqref{eq:gcdf}) with the parameters in \tabref{tab:gcdf}. 
    This rapid approach can help identify simulated datasets worth full Bayesian MCMC analysis.

    \item The detection fraction is 50\% at $\lnlike=4.88$ for a calibrated Bayes factor threshold of $\bfthresh=21$ and 5\% false alarm probability, when successful and false detection are distinguished by the frequency recovery criteria in Eqs.\ \eqref{eq:fo_width}--\eqref{eq:fo_delta}.
    Nondetections occur up to $\lnlike \sim 7.5$, above which all CWs are detected and most source parameters are well-recovered, with 68\% CI frequency posterior widths $\lesssim 0.5~\tr{nHz}$, 68\% CI strain amplitude posterior widths $\lesssim0.4~\tr{dex}$, and 68\% confidence sky localization areas $\lesssim 1000~\tr{deg}^2$. The $\mc$ recovery, which is extracted from frequency evolution, can be narrowly recovered for most high-mass, high-frequency detections, rarely for low-frequency detections.

    \item Source confusion between multiple competitive CWs can prevent successful detection of either of them, a problem that could be solved using trans-dimensional signal models. Meanwhile, source confusion between a CW and the background can create false detections when the CW model component captures red noise. The latter is distinguishable from the multiple CW scenario in that the falsely recovered frequency tends to be low, and the remaining GWB is captured by an unphysical model, with a suspiciously flat or even positive slope. This may be an effect of the CW+CURN approach and could be addressed by including spatial correlations in the GWB component of the MCMC model.

\end{enumerate}


\appendix 

\section{Semi-Analytic Model Parameter Spaces}\label{sec:appendix_pspace}
There are two parameter spaces defined in $\holodeck$ that we use for our binary population generation, titled $\classicuniform$ and $\asa$. 
$\classicuniform$ is described in detail in \citet{ng+23_astro}. This parameter space contains uninformative, broad priors, with six key varying parameters. 
\citet{gardiner+2024} also adopted this parameter space, taking the mean of each varying parameter as the fiducial model. 
Then, varying the parameters over the broad range allows the impact of individual model components on the GWB and CW detectability to be distinguishable. 
We take the same fiducial mean-parameter model as the starting point for our analysis, varying only the binary evolution parameters because they impact the GWB and CW most distinctly and are the least constrained by EM observations. 
This $\classicuniform$ mean-parameter model is effective in generating many low-$\snr$ realizations. However, to predict CW detection and characterization prospects in a higher-$\snr$ regime, we require a larger number of intermediate- to high-$\snr$ samples. These models should agree with prior astrophysical knowledge and the current GWB measurement to predict the most likely CW properties. Thus, we adopt the $\asa$ parameter space for the remaining samples, which draws all non-evolution parameters from normal distribution around literature-informed means and standard deviations. 
\response{
The second subset primarily uses the fiducial values of this model, except for some targeted variations to generate higher-$\snr$ realizations. 
However, in the third subset, drawing parameters from normal distributions around their literature values creates a wider variety of populations with varying GWBs, so that we can also incorporate constraints from the NG15 GWB measurement.
}

The GSMF and merger rate components are updated in $\asa$, while the $\mmbulge$ relation and binary evolution match those in the $\classicuniform$ SAM. The single Schechter GSMF is replaced with a double Schechter GSMF, the sum of two Schechter functions,
\begin{equation} \label{eq:gsmf_schechter}
    \frac{\partial\ndens}{\partial \log_{10}M} 
    = \ln(10)\Psi_1 \cdot \scale[\alpha_1 + 1]{M}{M_*} \exp \scale{-M}{M_*}
    + \ln(10)\Psi_2 \cdot \scale[\alpha_2 + 1]{M}{M_*}\exp \scale{-M}{M_*}
\end{equation}
The two components share a characteristic mass $M_*$, with constant, linear, and quadratic redshift dependence set by $ m_{\psi,z_0}$, $m_{\psi,z_1}$, and $m_{\psi,z_2}$ respectively, such that
\begin{equation} \label{eq:gsmf_mstar}
    \log_{10}(M_*) = m_{\psi,z_0} + m_{\psi,z_1} z + m_{\psi,z_2} z^2.
\end{equation}
The Schechter components $i=1$ and $i=2$ have distinct slopes set by $\alpha_1$ and $\alpha_2$ and distinct normalizations $\Psi_1$ and $\Psi_2$. Each normalization $\Psi_i$ is parameterized by a constant component $\psi_{i,z_0}$, linear component $\psi_{i,z_1}$, and quadratic component $\psi_{i,z_2}$ such that
\begin{equation} \label{eq:gsmf_norm}
    \log_{10} (\Psi_i) = \psi_{i,z_0} + \psi_{i,z_1} z + \psi_{i,z_2} z^2.
\end{equation}
The default parameters and normal distribution means are the best fits from \citet{leja+2020}; they are listed in \tabref{tab:asa_pspace}

An Illustris-motivated Galaxy Merger Rate (GMR) replaces the power-law Galaxy Merger Time and Galaxy Pair Fraction used in $\classicuniform$. The GMR $R$ is defined as the differential number of mergers as a function of $\mstartot$, $\qstar$, and $z$ per unit-time $t$, and mass ratio,  
\begin{equation} \label{eq:gmr_def}
    R(\mstartot, \qstar, z) = \diffp{N_\tr{mergers}(\mstartot,\qstar,z)}{{\qstar}{t}}
\end{equation}
The GMR is set by the fitting function defined in \citet{rodriguezgomez+2015} Table 1 as 
\begin{equation} \label{eq:gmr_func}
    R(\mstartot, \qstar, z) = 
    A(z) \scale[\alpha(z)]{\mstartot}{10^{10}M_\odot}
    \left[ 
    1+\scale[\delta(z)]{\mstartot}{2\times10^{11}\msol} 
    \right]
    \qstar^{\beta(z) + \gmrqgammam \log_{10}\scale{\mstartot}{10^{10}\msol}
    }
\end{equation}
where
\begin{eqnarray} \label{eq:gmr_params}
    A(z) = 10^{\gmrnormlog}\tr{Gyr}^{-1}(1+z)^{\gmrnormz}, \\
    \alpha(z) = \gmrmalphao (1+z)^{\gmrmalphaz},\\
    \beta (z) = \gmrqgammao (1+z)^{\gmrqgammaz}, \\
    \delta(z) = \gmrmdeltao (1+z)^\gmrmdeltaz.
\end{eqnarray}

The $\mmbulge$ relation to get SMBHB masses for the population of merged galaxies follows the same functional form as in Eq.\ (2) of \citet{ng+23_astro}
\begin{equation}
\label{eq:mmbulge_relation}
\log_{10}\! \lr{\mbh/\msol} = \mmbamp + \mmbplaw \log_{10}\!\scale{\mbulge}{10^{11} \, \msol} + \mathcal{N}\lr{0, \mmbscatter}.    
\end{equation}
with the amplitude normalization $\mmbamp$, power-law scaling $\mmbplaw$, and intrinsic scatter $\mmbscatter$ based on \citet{kh2013}.
$\classicuniform$ uses a constant bulge fraction $\bffbulge = 0.615$ to get $\mbulge$,
\begin{equation} 
    \mbulge = \bffbulge \mstartot.
\end{equation}
$\asa$ replaces the constant bulge fraction with a sigmoid dependence on stellar mass,
\begin{eqnarray} \label{eq:bulge_frac}
    \bffbulge(\mstartot < \bfmchar) &=& \bfflo + \frac{\bffhi - \bfflo}{1.0 + \left(\scale[-1]{\mstartot}{\bfmchar} - 1.0 \right)^\bfwidth} \\
    \bffbulge(\mstartot \geq \bfmchar) &\equiv& \bffhi
\end{eqnarray}
For stellar masses above $\bfmchar$, $\bffbulge=\bffhi$; then, below $\bfmchar$, $\bffbulge$ asymptotically approaches $\bfflo$ over a width set by $\bfwidth$, such that lower $\bfwidth$ corresponds to a steeper transition between $\bfflo$ and $\bffhi$.
While the constant $\bffbulge = 0.615$ is motivated by empirical observations of local galaxies \citep{bluck+2014, lang+2014}, little is known about bulge masses at higher redshifts in particular as they apply to merged galaxies. 
Thus, we set the broad uniform priors on $\bfflo$, $\bffhi$, and $\bfwidth$, listed in \tabref{tab:asa_pspace}.

Finally, the binary evolution matches that of $\classicuniform$  following Eqs.\ (6)-(8) in \citet{gardiner+2024}) which models the hardening rate $da/dt$ as the sum of a GW component and a phenomenological component, which represents all non-GW hardening mechanisms. To summarize the relevant parameters,
the phenomenological hardening rate $(da/dt)_\tr{phenom}$ is set by a small-separation hardening index $\hardnuinner$ and a large-separation hardening index $\hardnuouter$, with the two regimes separated by a critical break separation $\hardrchar$. 
$(da/dt)_\tr{phenom}$ is normalized to a fixed binary lifetime, $\tau_f$ representing the time between initial separation $\hardainit$ and coalescence.  The varying components, $\tau_f$, $\hardnuinner$, and $\hardnuouter$ are drawn from broad uniform priors reflecting the lack of EM constraints on binary evolution.


\begin{deluxetable}{cccc}
\label{tab:asa_pspace}
\tablecaption{$\asa$ Parameters for Our Semi-analytic Population Model}
\tablehead{
\colhead{Model Component} & \colhead{Symbol} & \colhead{Fiducial Value}  & \colhead{Astrophysical Priors}
} 
\startdata 
GSMF ($\gsmffunc$) 
& $\psi_{1,z_0}$ & -2.383 & $\normal{-2.383}{0.028}$ \\ 
& $\psi_{1, z_1}$ & -0.264 & $\normal{-0.264}{0.072}$ \\ 
& $\psi_{1,z_2}$ & -0.107 & $\normal{-0.107}{0.031}$ \\ 
& $\psi_{2,z_0}$ & -2.818 & $\normal{-2.818}{0.050}$ \\ 
& $\psi_{2,z_1}$ & -0.368 & $\normal{-0.368}{0.070}$
\\ 
& $\psi_{2,z_1}$ & +0.046 & $\normal{+0.046}{0.020}$
\\ 
& $m_{\psi,z_0}$ & +10.767 & $\normal{+10.767}{0.026}$    
\\ 
& $m_{\psi,z_1}$ & +0.124 & $\normal{+0.124}{0.045}$      
\\ 
& $m_{\psi,z_2}$ & -0.033 & $\normal{-0.033}{0.015}$      
\\ 
& $\alpha_1$ & -0.280 & $\normal{-0.280}{0.070}$           
\\ 
& $\alpha_2$ & -1.480 & $\normal{-1.480}{0.150}$     
\\ 
\hline
GMR ($R$) 
& $\gmrnormlog$ & -2.2287 & $\normal{-2.2287}{0.0045}$        
\\ 
& $\gmrnormz$ & +2.4644 & $\normal{+2.4644}{0.0128}$
\\ 
& $\gmrmalphao$ & +0.2241 & $\normal{+0.2241}{0.0038}$   
\\
& $\gmrmalphaz$ & -1.1759 & $\normal{-1.1759}{0.0316}$     
\\ 
& $\gmrmdeltao$ & +0.7668 & $\normal{+0.7668}{0.0202}$  
\\ 
& $\gmrmdeltaz$ & -0.4695 & $\normal{-0.4695}{0.0440}$ 
\\ 
& $\gmrqgammao$ & -1.2595 & $\normal{-1.2595}{0.0026}$   
\\ 
& $\gmrqgammaz$ & +0.0611 & $\normal{+0.0611}{0.0021}$
\\ 
& $\gmrqgammam$ & -0.0477 & $\normal{-0.0477}{0.0013}$    
\\ 
\hline
$\mmbulge$
& $\mmbamp$ & 8.69 & $\normal{8.69}{0.05}$ 
\\ 
& $\mmbplaw$ & 1.17 & $\normal{1.17}{0.08}$                   
\\ 
& $\mmbscatter$ & $0.28~\tr{dex}$ & $\normal{0.28, 0.05}$ dex
\\ 
\hline
Bulge Fraction 
& $\bfflo$ & 0.4 & $\uniform{0.1}{0.4}$
\\ 
& $\bffhi$ & 0.8 & $\uniform{0.6}{1.0}$
\\ 
& $\bfmchar$ & 11.0,              
\\ 
& $\bfwidth $ & 1.0 dex & $\uniform{0.5}{1.5}$ dex              
\\ 
\hline
Evolution ($\frac{da}{dt}$) 
& $\tau_f$ & 3.0 Gyr &   $\uniform{0.1}{11.0}$Gyr
\\ 
& $\hardainit$ & $1\times10^4~\tr{pc}$ & \nodata     
\\ 
& $\hardrchar$ & $10.0~\tr{pc}$ & $\uniform{2}{20} ~\tr{pc}$       
\\ 
& $\hardnuinner$ & -1.0 & $\uniform{-2.0}{0.0}$
\\ 
& $\hardnuouter$ & 0.0 & \nodata
\enddata
\tablecomments{The Galaxy Stellar Mass Function (GSMF) is described by Eqs.\ \eqref{eq:gsmf_schechter}--\eqref{eq:gsmf_norm}, the Galaxy Merger Rate (GMR) by Eqs.\ \eqref{eq:gmr_func}--\eqref{eq:gmr_params}, the $\mmbulge$ relation by Eq.\ \eqref{eq:mmbulge_relation}, the Bulge Fraction in Eq.\ \eqref{eq:bulge_frac}, and the Evolution model in \citet{gardiner+2024} Eqs.\ (6)--(8). 
}
\end{deluxetable}

\section{$\sigsig$ and $\sigdat$ Figures} \label{sec:appendix_othersnrs}

\begin{figure}
    \centering
    \includegraphics[width=0.49\linewidth]{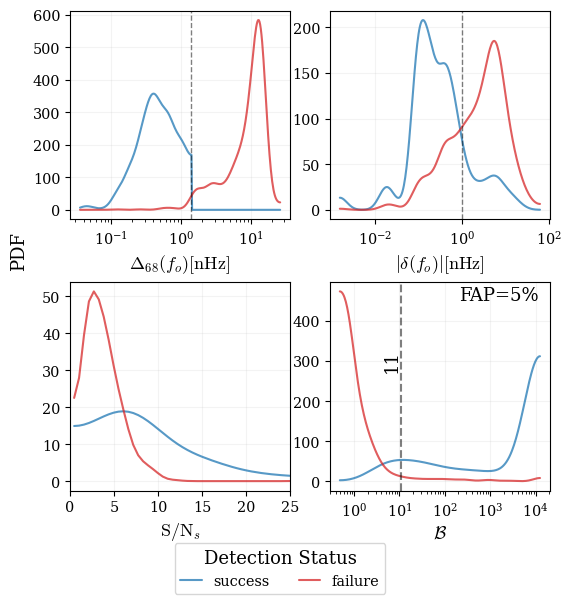}
    \includegraphics[width=0.49\linewidth]{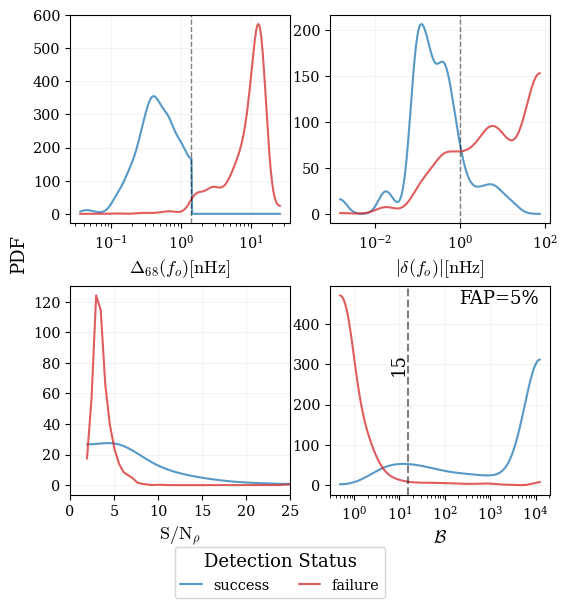}
    \caption{Same as \figref{fig:detected_dist} but using $\sigsig$ (left) and $\sigdat$ (right) to identify and rank the injected CWs.}
    \label{fig:detected_dist_appendix}
\end{figure}

\begin{figure}
    \centering
    \includegraphics[width=0.49\linewidth]{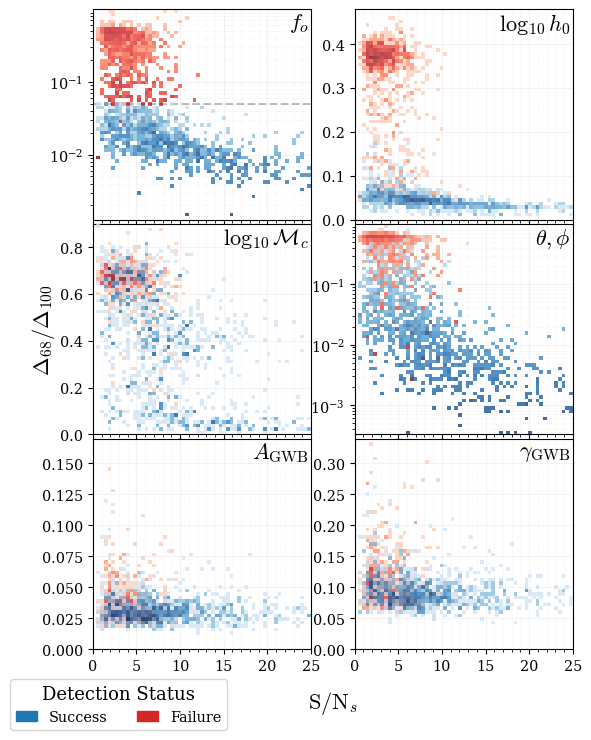}
    \includegraphics[width=0.49\linewidth]{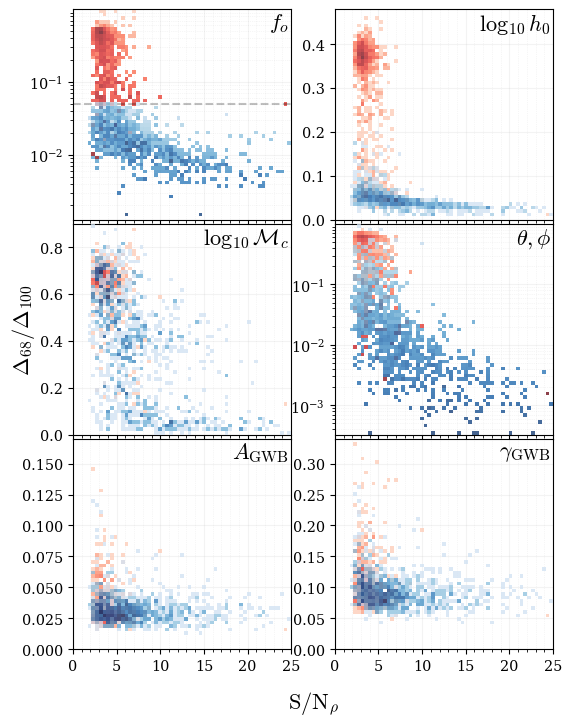}
    \caption{Same as \figref{fig:props_ci_vs_snr} but with the x-axis as $\sigsig$ (left) and $\sigdat$ (right).}
    \label{fig:props_ci_vs_snr_appendix}
\end{figure}

\begin{figure}
    \centering
    \includegraphics[width=0.49\linewidth]{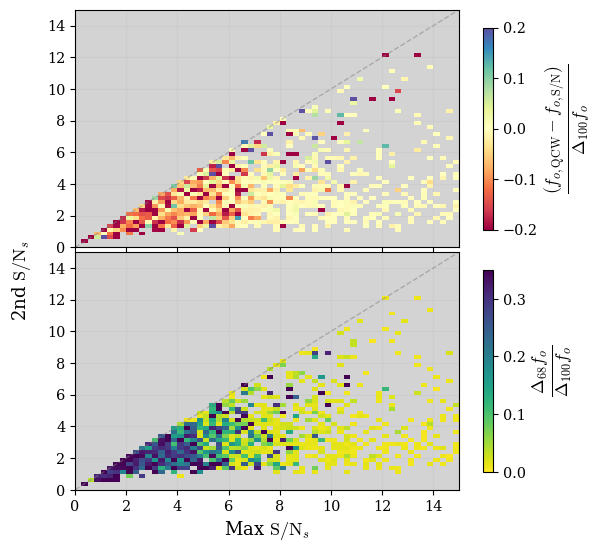}
    \includegraphics[width=0.49\linewidth]{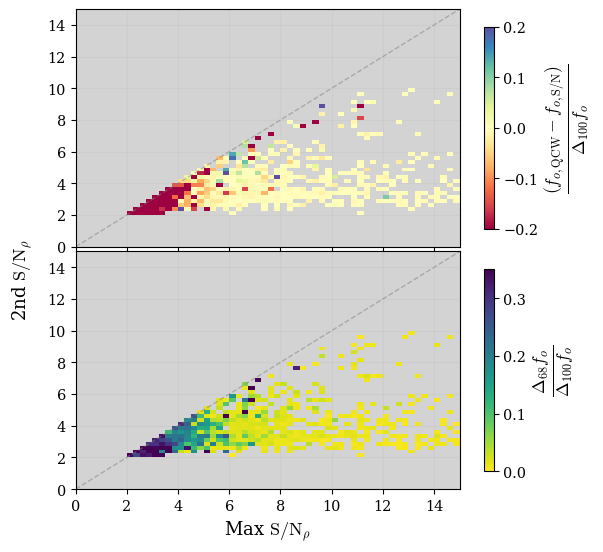}
    \caption{Same as \figref{fig:source_confusion} but for $\sigsig$ (left) and $\sigdat$ (right)}
    \label{fig:source_confusion_appendix}
\end{figure}

In \figref{fig:detected_dist_appendix}--\figref{fig:source_confusion_appendix} we present results based on $\sigsig$ (Eq.\ \eqref{eq:sigsig}) and $\sigdat$ (Eq.\ \eqref{eq:sigdat}). 

Comparing Figs. \ref{fig:detected_dist_appendix} and \ref{fig:detected_dist} shows that high $\sigsig$ appears in the null population more often than high $\sigdat$ or $\lnlike$, making it an unreliable estimator of detections. 
The $\sigdat$ and $\lnlike$ distributions are similar, but with $\sigdat$ having less frequency alignment in the null population. 
The noise contributions to $\sigdat$ make it more likely to be large at high frequencies, where white noise dominates. 
Meanwhile, nondetections of the MCMC tend towards lower frequencies, where the CW model component can capture red noise instead of true CW sources. 
This leads 
$|\delta(\fo)|$
in nondetection scenarios to be greater for the highest-$\sigdat$ injections than the highest-$\lnlike$ injections.


\section{Acknowledgements}
The authors are grateful to Polina Petrov and Stephen R. Taylor for helpful feedback during the drafting of this paper.
\response{The authors also thank the anonymous referee for providing constructive questions and comments, helping us improve the clarity and discussion of this paper.}
NJC appreciates the support of the NANOGrav NSF Physics Frontiers Center award No. 2020265. 
\software{
    \texttt{astropy} \citep{astropy},
    \texttt{cython} \citep{cython2011},
    \texttt{healpy} \citep{healpy},
    \texttt{holodeck} \citep{holodeck},
    \texttt{kalepy} \citep{kalepy},
    \texttt{ligo.skymap} \citep{ligoskymap_singer+price2016},
    \texttt{matplotlib} \citep{matplotlib2007},
    \texttt{numpy} \citep{numpy2011},
    \texttt{pint} \citep{pint2018, pint2024},
    \texttt{pta\_replicator} \citep{pta_replicator},
    \texttt{QuickCW} \citep{becsy+2022_quickcw},
    \texttt{scipy} \citep{2020SciPy-NMeth},
}


\bibliography{refs}{}
\bibliographystyle{aasjournal}



\end{document}